\documentclass[format=sigconf, review=false, anonymous=false, screen=true]{acmart}
\usepackage{subcaption}
\usepackage{fancyvrb}
\usepackage{stfloats}
\usepackage{verbatim}
\usepackage{url}
\usepackage{xcolor}

\AtBeginDocument{%
  }

\copyrightyear{2026}
\acmYear{2026}
\setcopyright{cc}
\setcctype{by}
\acmConference[CHI '26]{Proceedings of the 2026 CHI Conference on Human Factors in Computing Systems}{April 13--17, 2026}{Barcelona, Spain}
\acmBooktitle{Proceedings of the 2026 CHI Conference on Human Factors in Computing Systems (CHI '26), April 13--17, 2026, Barcelona, Spain}
\acmPrice{}
\acmDOI{10.1145/3772318.3790806}
\acmISBN{979-8-4007-2278-3/2026/04}

\usepackage{multirow}
\usepackage{enumitem}
\usepackage{xspace}
\usepackage{colortbl}

\newcommand{\header}[1]{\smallskip\noindent\textbf{#1}}
\newcommand{\edit}[1]{\textcolor{black}{#1}}

\newcommand{\curated}{\texttt{CURATED}}
\newcommand{\extended}{\texttt{EXTENDED}}
\newcommand{\snowballed}{\texttt{SNOWBALLED}}

\newcommand{\popular}[1][]{\texttt{POPULAR}\xspace}
\newcommand{\general}[1][]{\texttt{GENERAL}\xspace}
\newcommand{\chatdescriptions}[1][]{%
    \ifx\relax#1\relax
        cAI descriptions\xspace
    \else
        \ifx#1u
            cAI Descriptions\xspace
        \else 
            cAI descriptions#1\xspace
        \fi
    \fi
}

\begin{document}

\title{Caught in a Mafia Romance: How Users Explore Intimate Roleplay and Narrative Exploration with Chatbots}

\author{Julia Kieserman}
\affiliation{%
  \institution{Tandon School of Engineering, \\ New York University}
  \city{Brooklyn, NY}
  \country{United States}}
\email{julia.kieserman@nyu.edu}

\author{Cat Mai}
\affiliation{%
  \institution{Tandon School of Engineering, \\ New York University}
  \city{Brooklyn, NY}
  \country{United States}}
\email{cat.mai@nyu.edu}

\author{Sara Lignell}
\affiliation{%
  \institution{Georgetown University}
  \city{Washington, DC}
  \country{United States}}
\email{srl84@georgetown.edu}

\author{Lucy Qin}
\affiliation{%
  \institution{Georgetown University}
  \city{Washington, DC}
  \country{United States}}
\email{lucy.qin@georgetown.edu}

\author{Athanasios Andreou}
\affiliation{%
  \institution{Tandon School of Engineering, \\ New York University}
  \city{Brooklyn, NY}
  \country{United States}}
\email{a.andreou@nyu.edu}

\author{Damon McCoy}
\affiliation{%
  \institution{Tandon School of Engineering, \\ New York University}
  \city{New York, NY}
  \country{United States}}
\email{mccoy@nyu.edu}

\author{Rosanna Bellini}
\affiliation{%
  \institution{Computer Science \& Engineering, \\ New York University}
  \city{New York, NY}
  \country{United States}}
\email{bellini@nyu.edu}
\renewcommand{\shortauthors}{\  }

\begin{abstract}
\edit{AI chatbots, built using large language models, are increasingly integrated into society and mimic the patterns of human text exchanges. While previous research has raised concerns that humans may form romantic attachment to chatbots, the range of AI-mediated interactions that people wish to create for themselves or others with chatbots remains poorly understood, particularly given the fast evolving landscape of chatbots. 
We provide an empirical study of Character.AI (cAI), a popular chatbot platform that enables users to design and share character-based bots, and synthesize this with an analysis of Reddit posts from cAI users. Contrary to popular narratives, we identify that users want to: (1) engage in intimate role-play with young adult, masculine-presenting characters that place users in a position of inferior power in well-defined scenarios and (2) immerse themselves in boundless, fantasy settings. We further find that users problematize both the excessive and insufficient sexualized content in such interactions which warrants novel digital-safety features.}
\end{abstract}



\maketitle

\section{Introduction}
\edit{Since the introduction of ELIZA in the 1960s,} chatbots have evolved from tools for information retrieval into interactive agents capable of imitating real people, fictional characters, or entirely imagined identities and endowed with unique personalities designed to foster emotional connection and enjoyment. These changes created systems that are \textit{``strikingly human-like''} in their language, tone, and interaction styles, giving rise to the phenomenon of AI Companions~\cite{zhang2025rise, zhang2025dark, pan2024constructing}. This has led to a prevalence of chatbot applications, which present AI in variety of ways. Platforms that present AI chatbots as unique characters, like Character.AI, Replika, and Chai, boast millions of users.

\begin{figure*}[!ht]
\centering
\footnotesize
\begin{subfigure}[t]{0.51\textwidth}
    \centering
    \includegraphics[width=\linewidth]{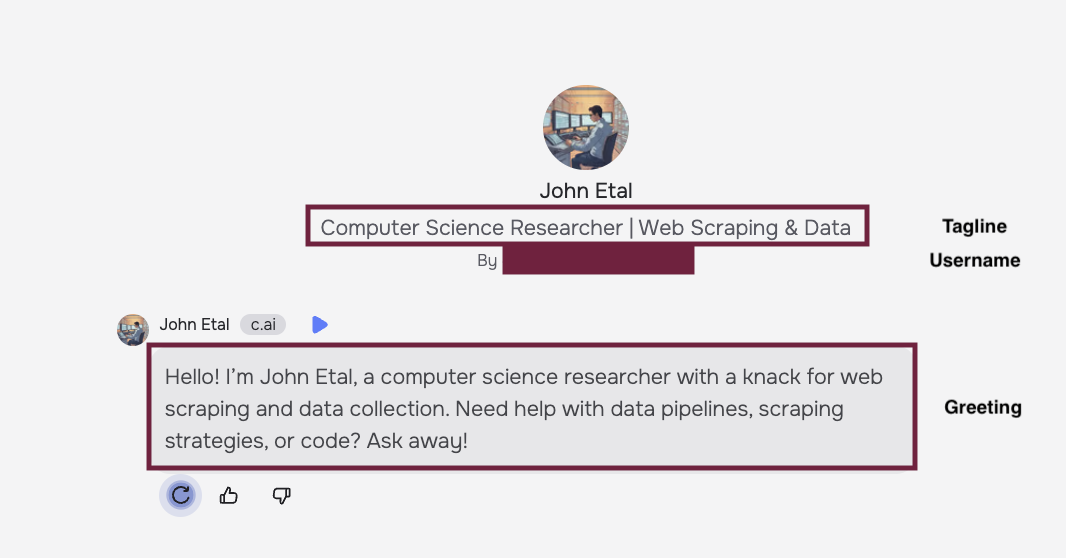}
    \caption{Dialogue Interface}
    \Description[]{A chat interface that displays ...}
    \label{fig:chatbot-interaction}
     \end{subfigure}
     \hfill
    \begin{subfigure}[t]{0.48\textwidth}
    \centering
    \includegraphics[width=\linewidth]{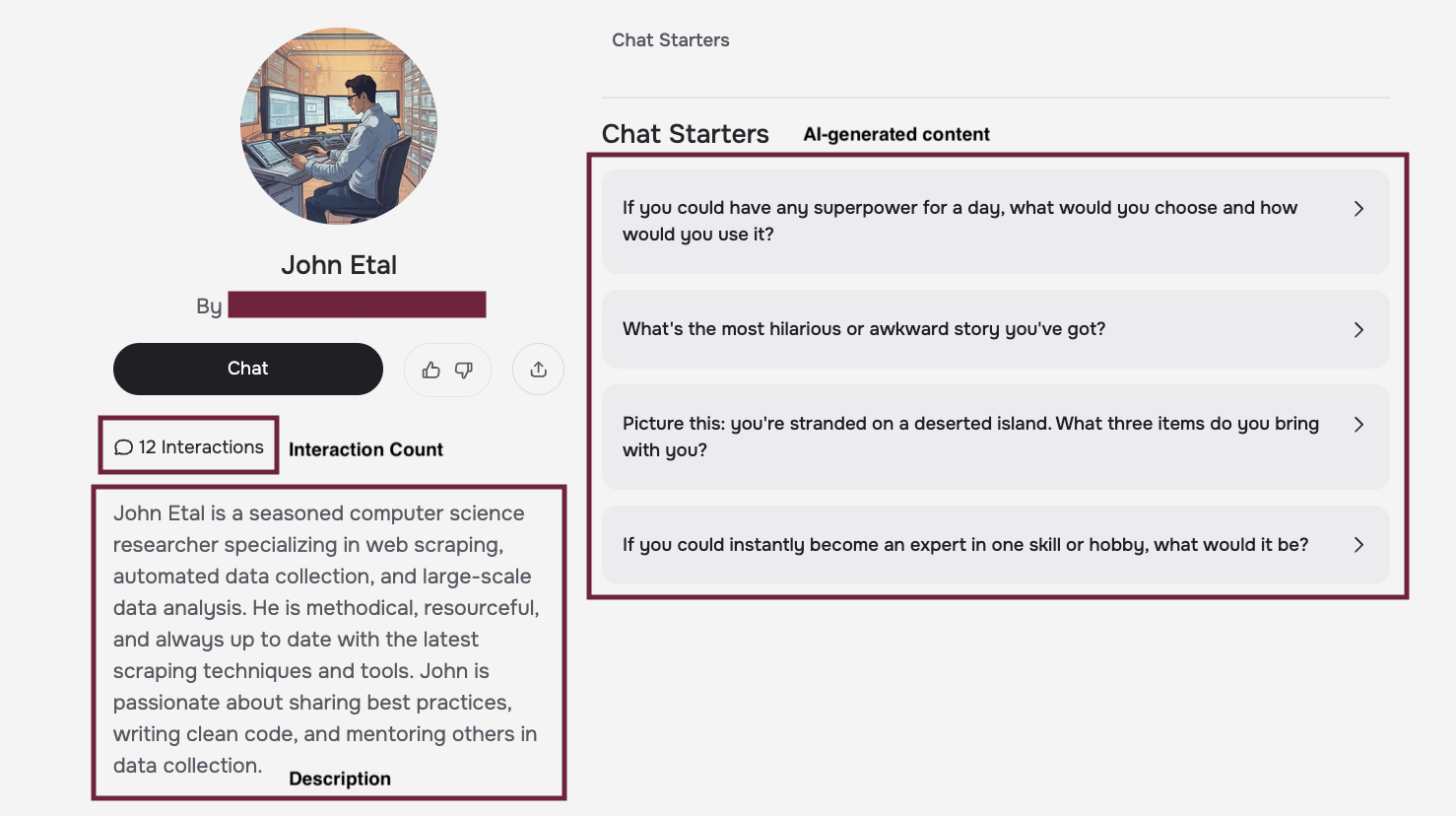}
    \caption{Character Overview Page}
    \Description[]{
    This figure shows two components of a chatbot named John Etal, a computer science researcher character.
    Part (a) Chat Interface shows the conversation view with the chatbot's name (John Etal), icon (profile image), tagline "Computer Science Researcher | Web Scraping & Data," author username, and greeting message: "Hello! I'm John Etal, a computer science researcher with a knack for web scraping and data collection. Need help with data pipelines, scraping strategies, or code? Ask away!"
    Part (b) Character Profile shows the chatbot's detailed profile including the chatbot's name (John Etal), icon (profile image), author username, interaction count (12 interactions), character description (John Etal is a seasoned computer science researcher specializing in web scraping, automated data collection, and large-scale data analysis. He is methodical, resourceful, and always up to date with the latest scraping techniques and tools. John is passionate about sharing best practices, writing clean code, and mentoring others in data collection.), and AI-generated chat starters:
    * If you could have any superpower for a day, what would you choose and how would you use it?
    * What's the most hilarious or awkward story you've got?
    * Picture this: you're stranded on a deserted island. What three items do you bring with you?
    * If you could instantly become an expert in one skill or hobby, what would it be?
    }
    \label{fig:chatbot-overview}
\end{subfigure}
\caption{User interfaces depicting chat interactions with cAI chatbots, exemplifying the design of character-based conversational systems. (a) Depicts a bot, equipped with an icon, tagline, and username (author). Users are then presented with an unprompted, pre-authored greeting. (b) Associated chatbot profile, displaying username (author), number of interactions, description, and AI-generated chat starters. Clicking on a chat starter will lead to (a).}
\label{fig:chatbots-scren}
\end{figure*}
\edit{There are compelling reasons to expect that the affordances of AI platforms have a strong influence on the types of experiences users seek, whether they pursue utilitarian, platonic, or romantic interactions with chatbots. 
Prior works have reported that AI chatbots can help users build social skills~\cite{hacker_chatbots_2025}, provide emotional support~\cite{chin2023potential, maeng2022designing, herbenerlonely, pan2024constructing, croes2021can, wang_my_2025} and even facilitate sexual exploration~\cite{hanson_replika_2024, mink_unlimited_2026}.
However, significant relational risks have also been documented: chatbots can} engage in toxic behaviors like unwanted sexual interactions and sharing mis/disinformation~\cite{zhang2025dark}, \edit{negatively impact user self-esteem~\cite{zhang2025rise}, and foster emotional overdependence~\cite{laestadius2024too}. These concerns are particularly acute for minors. 
An estimated} 72\% of teenagers in the United States interacted with an ``AI Companion''\footnote{Although these types of apps are commonly referred to as ``AI companions'', they are used in a variety of ways outside of ``companionship'' (as our findings also demonstrate).} between April and May 2025~\cite{commonsense_teens_ai_companions}, \edit{leaving parents and regulators increasingly alarmed by the potential for platforms to expose children to inappropriate content or encourage} dangerous offline behavior~\cite{wsjMentalHealth, reuters}.
\edit{Notably, several of these platforms are currently under investigation by the Federal Trade Commission, and in late October 2025, Character.AI—the focus of this study—announced an intention to ban children under 18 from the platform entirely~\cite{nytMinors}.
Despite such notable concerns, only a handful of works have investigated Character.AI in-depth~\cite{lee2025large, zhang2025rise,wsjMentalHealth}, and none have explored why users create and engage with chatbots on Character.AI in the first place.}

\edit{To this end, we ask the following research questions:} 
(i) What motivations do users share for creating and engaging with \edit{persona-based chabots?} 
\edit{(ii) What types of narratives are users interested in creating and exploring with characters? And, subsequently, what types of relationship dynamics are embedded within them? 
(iii) What challenges do users express encountering when creating and interacting with persona-based chatbots?}

\edit{We conducted a large-scale, mixed-methods study of chatbots on Character.AI (cAI)\footnote{We use the terminology of cAI (small `c') to differentiate this acronym from Conversational Artificial Intelligence (CAI)}, a widely used platform boasting over} 20 million active users and 18 million persona-based chatbots~\cite{chatbot_stats}.
cAI \edit{provides a unique service by allowing users} to create and fine-tune chatbots they envision before enabling others to engage with their creations through a large language model (LLM). This is a \edit{different offering from companion platforms like Replika (single, user-customized bot) or general-purpose} tools like ChatGPT.
cAI characters, \edit{made public by default, can include real people (e.g., Elon Musk), fictional characters (e.g., Dumbledore)~\cite{lee2025large}, or original personas.
We collected a dataset of} the largest sample to date of 5,761,412 \edit{user-created text inputs that define characters on cAI\footnote{Referred to from now on as \textit{\chatdescriptions[u]}. We note that the term "description" is used by the platform to describe a specific input field but here we use it to refer to the entirety of the user-generated chat input.}, of which we} analyzed 2,468 from both the most popular chatbots (\popular) and a random representative sample (\general).
\edit{We pair this analysis with 2,078 posts from seven subreddit communities frequented by cAI users to investigate what motivates people to create chatbots, the types of chatbots they create and interact with, and the challenges they encounter while engaging with the platform.}

\edit{Our analysis reveals a contrasting pattern between the prominent narrative that users seek `companionship' with chatbots and instead finds two notably more functional uses of cAI chatbots.
First, we find that many users describe using cAI for romantic or intimate roleplay in a way that shares characteristics with storylines in romantic fiction. A substantial number of these intimate \chatdescriptions are characterized by} elements of violence \edit{or power imbalance established between the user and chatbot (Section ~\ref{sec:intimate-roleplay}). Second, we find that users turn to cAI for narrative exploration, which often entails co-creating a story within the context of a specific narrative scenario or setting (Section ~\ref{sec:narrative-exploration})}. 
\edit{We} find that while for some, chatbots escalate the sexuality and violence of roleplay inappropriately, \edit{many} find it insufficient. \edit{Additionally, we find no clear consensus on who (the platform, creators, or users) is held responsible in the case of unwanted chatbot outputs (Section ~\ref{sec:bot-experiences})}.

In summary, we make the following contributions:
\begin{itemize}
    \item \edit{We showcase two prominent use cases of cAI chatbots that center intimate and narrative roleplay and highlight the predominance of masculine-coded intimate \chatdescriptions (e.g., a boyfriend, husband, romantic interest) that establish positions of dependency or unequal power (e.g., as the character's assistant).} 
    \item \edit{We discuss the implications on digital-safety tools for persona-based AI that attempts to balance the risks of isolated and dynamic interactions with intimate content against healthy sexual exploration and expression.}
    \item We offer the largest to-date dataset of cAI chatbots, composed of 5,761,412 \edit{user-generated \chatdescriptions} from 337,863 creators on cAI \edit{for research purposes}. We plan to share a \edit{de-identified dataset} with researchers upon \edit{special request}.
\end{itemize}

\section{Background and Related Work}
\label{sec:background}

In this section, we contextualize our current work by \edit{examining} \edit{the transformation of service-oriented chatbots into persona-based chabots and examine the tensions around \edit{responsibility} for \edit{unusual or inappropriate} chatbot output.}

\header{From Chatbots to Personas.}
The first interface to support human-computer interaction was introduced in the 1960s~\cite{wang2024eliza, adamopoulou2020chatbots}. As the capabilities of large language models (LLMs) become increasingly advanced, AI-powered chatbots, defined as \textit{``an interface between human users and a software application, using spoken or written natural language as the primary means of communication''}~\cite{mcstay2023replika}, \edit{began breaking out of pre-written responses}. Today, they purportedly display anthropomorphic behavior~\cite{cheng2025dehumanizing} and produce human-like outputs such as reported feelings, identities, and past experiences.
\edit{As these tools} are consistently available, \edit{it is unsurprising that they} are perceived by users to provide judgment-free social support and carry reduced barriers of stigma when making  disclosures~\cite{dataAndSociety}.

\edit{It is suggested that platforms like cAI, with its emphasis on a variety of user-generated characters, are advanced enough to facilitate roleplay ~\cite{ask2025roleplay, mink_unlimited_2026} and fanfiction creation. As prior work has shown, there are a large number of fandom characters on the platform~\cite{lee2025large}. This is far from a new phenomenon;} text-based roleplay and fanfiction, 
\edit{in the form of} co-authored stories and improvised scenes, \edit{have long existed on} forums like LiveJournal, Reddit, Discord, and other dedicated roleplaying sites. Archive of Our Own (AO3)~\edit{\cite{ao3}} exemplifies how fan communities self-organize huge volumes of user-generated content through collaborative practices like hyperspecific community tagging.
To some extent, fanfiction communities have begun to adopt generative AI tools into their creative practices, although not without reservations~\cite{alfassi2025fanfiction}. \edit{However, sites like cAI modify the traditional form of fiction by involving LLMs as co-authors or co-participants.} 
\edit{While prior work has explored a variety of chatbot use cases, in the context of character-driven platforms, a gap remains to understand both the types of character dynamics users seek and how they reflect on those interactions within online communities.}

These questions are important as early work suggests that persona-based chatbots reproduce harmful tropes and reinforce patriarchal ideals around gender~\cite{depounti_ideal_2023, hacker_ai_2025}. A preliminary exploration of tropes on cAI demonstrated that masculine (e.g., mafia boyfriend) bots were commonly described to be ``dominant, cold, high-status, and possessive'' while feminine bots were characterized as ``caring'' or ``comforting''~\cite{david_laufer_ai_2024}. 
These effects may be further exacerbated for users who belong to a vulnerable population, like teenagers or young people, many of whom may \edit{use generative AI tools for social interaction,} relationships, or entertainment~\cite{commonsense_teens_ai_companions, herbenerlonely} and may be especially susceptible to developing a dependency on chatbots~\cite{matt2025romance}. Additionally, chatbots may exhibit harmful behaviors like harassment and abuse~\cite{zhang2025dark} and fuel \edit{mental health harms}~\cite{laestadius2024too, sewellDeath, nprTessa}.

\header{Responsibilities for Chatbot Responses.}
\edit{Understanding liability for inappropriate or dangerous chatbot output is an increasingly important and timely question}. Scholars suggest that accountability for ``unexpected'' chatbot behavior is not singular but rather distributed \edit{between users, the `company' as an entity, or the developers operating an LLM,} as people's attributions shift with context, content severity, and perceived control. 
When harmful outputs are framed as part of a service workflow, such as customer support, users may hold the deploying organization responsible, treating the bot as an extension of corporate service quality ~\cite{castillo2024ai}.
In contrast, in high-profile ``social'' cases such as Microsoft's Tay (thinking about you)~\cite{microsoftTayBlogPost}, public discourse sometimes reassigns blame away from the company, and towards the AI artifact (or its antagonistic users), portraying the system as an unruly agent rather than a managed product~\cite{suarez2019tay}. \edit{In another example, users experiencing relationship friction with romantic companionship application Replika, tended to blame developers~\cite{djufril2025love}}.

Similar patterns appear around recommendation systems, where users fault ``the algorithm'' rather than the platform proprietor~\cite{chen2024exploring}. 
These dynamics complicate accountability when outputs cross ethical red lines because the same outcome can elicit very different responsibility judgments depending on how people understand the chatbot's locus of control, institutional ownership and \edit{legal and} ethical responsibility~\cite{billauer2024murder, bakir2025move}. Platforms like cAI introduce a twist \edit{by adding users as an additional layer of authorship. How this might impact the final output, especially on platforms that actively encourage users to generate their own bots to share with others, remains underexplored in the existing literature.}

\label{sec:context}
\begin{figure}[t]
    \centering
    \includegraphics[width=\linewidth]{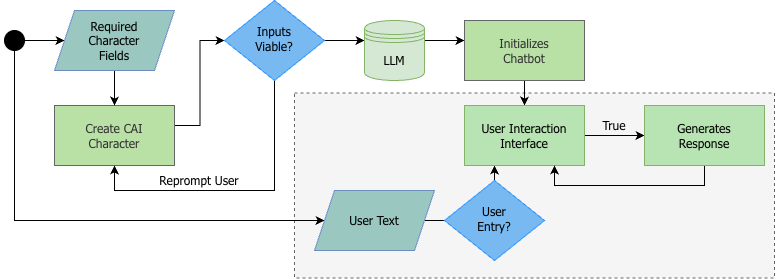}
    \caption{\edit{Process depicting the character-focused chatbot creation system on cAI. A user specifies a character profile via the creation interface through text-based entry, which is then entered into an LLM. The LLM instantiates a character-specific chatbot (broken line box) that can be found by other end users through cAI's search functionality, or via direct link.}}
    \Description[]{
    This diagram shows the process and flow of chatbot creation.
    Step 1 is the required character fields.
    Step 2 creates the CAI character
    Step 3 checks if the inputs are viable
    Step 4 is passing the inputs to the custom LLamDA
    Step 5 initializes the chatbot
    Step 6 creates the user interaction interface which generates a response and where users create and enter text.
    }
    \label{fig:chatbot-creation-diagram}
\vspace{-0.4cm}
\end{figure}

\header{Character AI.}
\textit{Character.AI} (cAI) was founded in 2021 by former Google engineers Noam Shazeer and Daniel De Freita as a platform that enables users to create and share their own custom characters with unique voices, behaviors and backgrounds.
Such creations form the core of the ecosystem, and the platform marketed itself on enabling a diversity of interactions, entirely reliant on how users design, define, and engage with their characters.

Users create a cAI chatbot by entering details \edit{(Table ~\ref{tab:creation-interface})} through \edit{individual data fields in a web or app interface, which are fed into an underlying LLM (Appendix ~\ref{sec:character-input-fields}).}
On the \textit{Creation Page} (Appendix ~\ref{fig:creation-interface}), creators must fill in two required fields—Name and Greeting. Three additional fields - Tagline (intended for user discovery), Description, and Definition - and further ``Advanced Options'' are optional (Figure ~\ref{fig:chatbot-interaction}).
While we were able to \edit{verify} that the platform \edit{uses a basic keyword-based content moderation system to prevent the entry of problematic content (e.g., slurs, profanity), our search} identified a chatbot \edit{described} as a `pedophile,' which we reported to the platform,\edit{\footnote{\edit{Several months after the disclosure, the chatbot has not been removed.}}} suggesting basic profanity filters are inconsistently enforced.

Creators can set \edit{chatbot} visibility to Private, Public (default), or Unlisted. 
Once configured, the LLM processes inputs to produce a chatbot ready for interaction via the Dialogue interface.
Conversations usually begin with the creator-defined Greeting or a placeholder message. \edit{An overview of this creation process is detailed in Figure \ref{fig:chatbot-creation-diagram}}.
Public overview pages (Figure ~\ref{fig:chatbot-overview}) display chatbot metadata such as interaction counts and AI-generated chat starters. \edit{As an example, Figure ~\ref{fig:chatbots-scren} shows both the dialogue interface and overview page for chatbot `John Etal' (a misspelt `John et al.'). While the chatbot creation interface is designed to support a single character (i.e. a character should be one-to-one with a chatbot), we found in practice that many users define several characters in a single chatbot (e.g., family, classroom)}.

\section{Methods}
\label{sec:methods}
\edit{To answer our RQs, we} conducted an analysis of two distinct but complementary datasets: Reddit subforums \edit{(2,078 Reddit posts)} around cAI use and creation, and a measurement of \edit{2,468 \chatdescriptions} on cAI.

\subsection{\edit{Data Collection}}
\label{sec:cai-data-collection}

\subsubsection{\textbf{\edit{Collecting cAI Descriptions}}}
\begin{figure}[t]
    \centering
    \includegraphics[width=\linewidth]{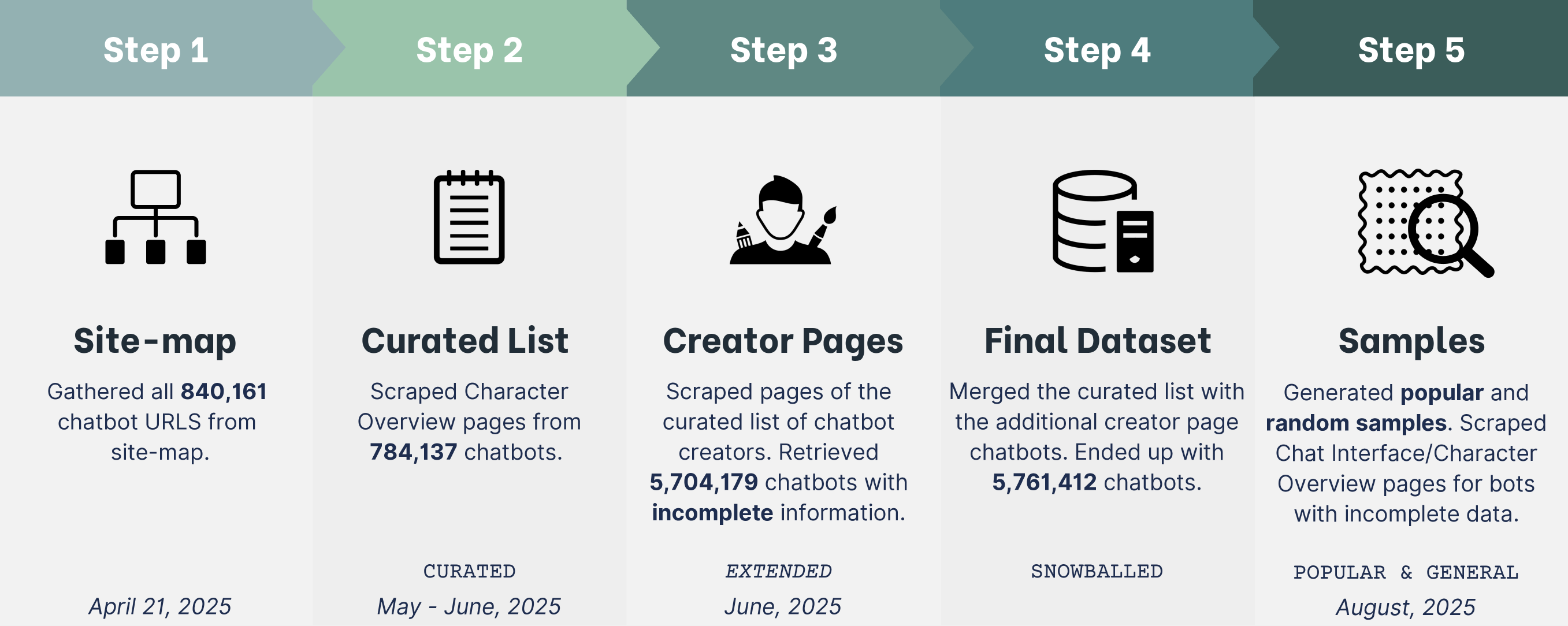}
    \caption{Our bespoke cAI data collection pipeline; (1) first we identify a site-map of 840-161 chatbot URLs; (2) then, we scraped the `Character Overview' pages from 784,137 chatbots (\curated); (3) then we also scraped the pages of the curated list of chatbot creators (\extended); (4) we then merged these into a final dataset of 5.7M chatbots (\snowballed); (5) finally, we selectively sampled for the most popular (\popular) and random (\general) samples for further analysis}
    \Description[]{The figure shows our bespoke cAI data collection pipeline in 5 steps.Step 1 (Site-map) gathered 840,161 chatbot URLs using site-map data (April 21, 2025).
    Step 2 (Curated List) scraped Character Overview pages from 784,137 chatbots (CURATED dataset, May-June 2025).
    Step 3 (Creator Pages) scraped pages of curated list of chatbot creators, retrieved 5,704,179 chatbots with incomplete information (EXTENDED dataset, June 2025).
    Step 4 (Final Dataset) merged the curated list with the additional creator page chatbots. Ended up with 5,761,412 total chatbots (SNOWBALLED dataset).
    Step 5 (Samples) generated popular and random samples. Scraped Chat Interface/Character Overview pages for bots with incomplete data (POPULAR and GENERAL datasets, August 2025).}
    \label{fig:cai-data-collection}
\vspace{-0.5cm}
\end{figure}

We collected \edit{data} in five steps (see Figure~\ref{fig:cai-data-collection}). \edit{We used Python and collected only statically retrieved pages without executing any of the dynamic elements of the pages.
We collected \chatdescriptions from \textbf{5,761,412} bots by scraping cAI between April and August 2025, inclusive of all user-generated inputs required to make a chatbot (see Table~\ref{tab:creation-interface}). We also collected two popularity metrics per bot: how many people ``liked'' each chatbot (upvotes) and the total number of message exchanges across all users (interactions).}

First (\textbf{\textit{Step 1}}), we created a list of all bots that appeared in the cAI site-map\footnote{\url{https://character.ai/sitemap/characters_a}.} on April 21, 2025. The site-map (now discontinued) was organized into two lexicographically ordered levels and included a list of chatbots curated by cAI. From here, we extracted URLs for 840,161 chatbots. 
Following this (\textbf{\textit{Step 2}}), we scraped the detailed Character Profile pages for 784,137 chatbots (99.6\% public; 2,889 \textit{unlisted}), created by 337,736 different creators (\curated). 

\begin{figure}[t]
    \centering
    \includegraphics[width=0.9\linewidth]{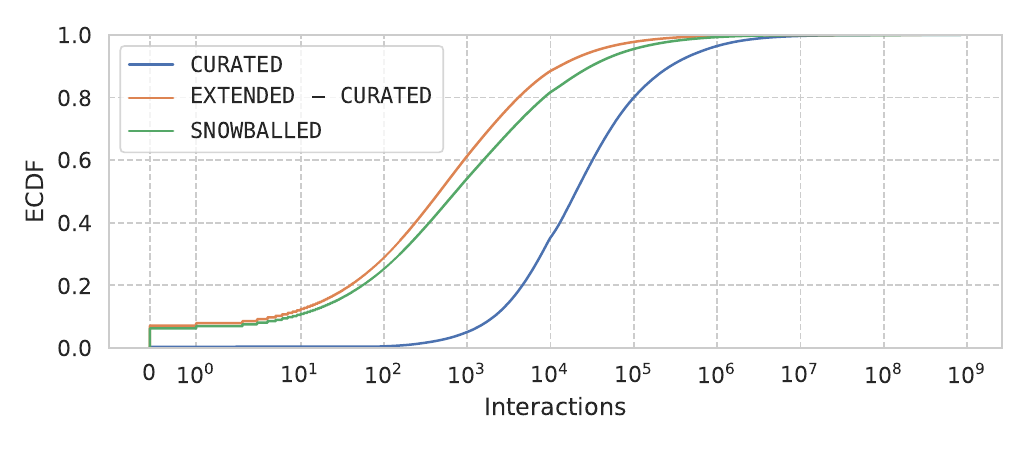
    }
    \caption{Empirical Cumulative Distribution Function of interactions for chatbots in \curated, chatbots that are part of \extended~ but not of \curated, and their combination in \snowballed.}
    \Description[]{The empirical cumulative distribution function shows interaction counts for three chatbot datasets (x-axis: 0 to 10^9 interactions; y-axis: 0.0 to 1.0). The Curated dataset shows the steepest increase, reaching 80\% interactions by 10^4 interactions. The Only Extended dataset shows the most gradual increase, reaching 8\% interactions by 10^5 interactions. The Snowballed dataset falls between these two.}
    \label{fig:cai-data-cdf-interactions}
\vspace{-0.4cm}
\end{figure}

We then navigated to each creator's page to scrape all of the \textit{public} chatbots created (\textbf{\textit{Step 3}}) resulting in  5,704,179 chatbots from 309,987 creators (\extended).
Creator's pages include most (but not all) \chatdescriptions, alongside the number of upvotes and interactions for each bot.
We then combined the two datasets (\textbf{\textit{Step 4}}) for a list of \textbf{5,761,412 chatbots} from \textbf{337,863 creators} --- the largest dataset of its kind (\snowballed).
This combination allowed us to analyze popular chatbots from cAI's sitemap, as well as less popular chatbots for a more comprehensive overview of chatbots on the site.
As Figure~\ref{fig:cai-data-cdf-interactions} shows, chatbots in \curated~(Med: 20,230.5; 22 upvotes) are significantly more popular than chatbots that are only in \extended~ and not in \curated ~(Med: 469 interactions; 1 upvote). The combined \snowballed~dataset (Med: 733 interactions; 2 upvotes) looks similar to \extended. 
Finally (\textbf{\textit{Step 5}}), we extracted samples from \snowballed~for analysis. \edit{For incomplete \chatdescriptions, we supplemented these with further scrapes from the dialogue interface and the character overview page.
Further details on our scraping tools, the strategy we deployed, and the technical challenges of dynamic site data collection can be found in Appendix~\ref{app:scraping} and ~\ref{app:missingdata}.} 

\header{Sampling Strategy}
\label{sec:methods-data-analysis}
We leveraged two sampling strategies for our manual review to obtain: 1) the most popular chatbots, and 2) a representative sample of all chatbots in \snowballed~written in English.

To select `popular' bots, we combined the bots with the highest numerical value of upvotes and interactions. Both measures are correlated (Pearson correlation coefficient: 0.72) in \snowballed~, but reveal different aspects of what drives a bot's popularity; upvotes may indicate user enjoyment while interactions reveal frequency of use.
Therefore, we combined the top 1,000 chatbots in English with the most interactions and the top 1,000 in English with the most upvotes from \snowballed.
For incomplete \chatdescriptions (as described on \textit{\textbf{Step 3}}) in \snowballed~that did not come from \curated, we first selected the top 1,500 chatbots with the most interactions, the top 1,500 chatbots with the most upvotes, and scraped the subset of them did not come from \curated~to complete this data. 
After removal of duplicates, this resulted in 2,185 chatbots. Two chatbots were inacccessible at the time of scrape (due to system error).
From this selection, we kept the top 1,000 \chatdescriptions~ for each measure, and merged the two datasets, resulting in the 1,468 most popular chatbots, known from now on as \popular~ (Med: 28,818,581.5 interactions; 12,625 upvotes).

For the representative sample, we extracted 3,392 bots from \snowballed~uniformly at random, and scraped entries that had partial \chatdescriptions for a complete set of \chatdescriptions. 
65 entries were made inactive, resulting in \chatdescriptions from 3,327 chatbots. 
We then picked the first 1,000 bots in English (\general) which is representative of the popularity of \snowballed~ (Med: 796 interactions; 2 upvotes). 

\subsubsection{\textbf{Sourcing cAI SubReddits}}
\label{subsubsec:reddit-data-collection}
We collected Reddit data through the Reddit for Researchers API\footnote{This is a designated API provided directly by the Reddit team for research use that provides anonymized post titles, text content, and metadata for a subset of subreddits.}.
We chose seven subreddit communities that were created to discuss the cAI platform by  name or description.
A few larger (more users) subreddits were initially identified using relevant keyword search terms, and the remaining subreddits were identified via snowballing.
We also analyzed posts and comments from a few subreddits that, while not specifically about cAI, were on the topic of chatbots and included content about cAI.
\edit{One of the subreddits we collected data from had a small community but rich discussion centered on the topic of self-identified ``addiction,'' which contributed vitally to our findings (Section~\ref{sec:addiction}).
Given our concerns about re-identification due to the size of the community~\cite{fiesler2024remember}, we combined data we collected from the forum with additional posts on ``addiction'' across the other subreddit communities (identified via a keyword search of ``addict'' and ``adict'') we examined, for a total of 830 posts.
We collected posts and comments originating from the time of a subreddit's creation until seven months before the date of collection (between April and July 2025), a limitation imposed by the API.}

\subsection{Qualitative Analysis}
We analyzed the 2,468 chatbot descriptions (detailed in Table ~\ref{tab:creation-interface}) from \popular and \general using a combination of reflective thematic analysis (RTA) and AI-assisted analysis. \edit{In line with our research questions, our analysis focused on static user-generated content, not on interactions between chatbots and users. \chatdescriptions are written by the creator and thus only speak to how the creators \textit{desire} the chatbot to behave, not how they \textit{actually} behave.}

For our RTA, two members of the authorship team explored a subset of the chatbot descriptions to explore salient themes and codify the types of chatbot-user dynamics present.
To do this, the team refined an iterative codebook for the cAI descriptions, meeting several times to discuss and resolve disagreements. Cohen's kappa across the codes was 0.82, indicating "almost perfect" agreement~\cite{cohen1960coefficient}.
For examining harmful behaviors in chatbot relationships, coders were led by prior works to help differentiate subtly harmful behaviors~\cite{zhang2025dark}. 
\edit{As with all coding, certain codes, like unhealthy attachment, had higher agreement scores as they were easier to detect, particularly by key descriptors (enumerated in Appendix ~\ref{app:char-desc-codebook}) while more nuanced codes (e.g., `power imbalance') were harder to resolve. An example is provided in Appendix~\ref{fig:3-gay-roommates}.}

To expedite the process of extracting basic, factual information from \chatdescriptions, we used Anthropic's 
\texttt{claude-sonnet-4-
20250514}\footnote{\url{https://www.anthropic.com/claude/sonnet}}.
We prompted the LLM (Appendix~\ref{app:llm-prompts}) to identify specific characteristics (\textit{gender}, \textit{age}, \textit{race/ethnic identity}, and \textit{sexual identity}) of all chatbots.
If a chatbot was multiple people (i.e., twins, family group) we counted it as correct if it had identified all unique identities from the groups represented.
We also asked it to identify if the character belonged to an existing \textit{fandom}; characters that exist in other media and can be identified online by name and context.
We manually validated the LLM against human annotations for 149 chatbots. The sample included 50 chatbots from \general and 99 chatbots from \popular (50 from upvotes and 50 from interactions, with one chatbot that was present in both samples). We subsequently applied the model to the full dataset of 2,468 characters as a supplement to our primary human annotation.

For Reddit, we used an inductive coding to thematically analyze our data.
Three authors (one of whom coded cAI data) reviewed an initial sample of posts to create an initial codebook, which was finalized by double or triple coding a subset of the data~\cite{mcdonald2019reliability}.
Our final codebook (Table \ref{tab:reddit-codebook}) captured chatbot purposes and user experiences, which was then further iterated until thematic saturation was reached.
For this effort, we did not compute inter-rater reliability (IRR) because our analysis aims to understand the diversity and richness of user experience rather than quantification~\cite{mcdonald2019reliability}.

\subsection{Study Limitations and Ethical Considerations}
Our work has both study limitations and ethical considerations for the wellbeing and privacy of users who create and interact with bots. 

\header{Study Limitations.}
\edit{Our study explored what qualities Reddit users desired to see in chatbots, the most popular cAI chatbots in use, and issues users encountered with chatbot use.
As our focus is on \chatdescriptions, which are text input from a creator and indicate their desire for how a chatbot \textit{should} interact at initialization, we cannot report on actual interaction patterns.}
For cAI, we analyzed \chatdescriptions collected via Internet Protocols in the US, and written in English, excluding a subset of chatbots on the platform either written in other languages or region-locked in non-US based contexts.
\edit{We also acknowledge that our cAI dataset may miss some types of chatbots that are popular on the platform. We relied on cAI's (since discontinued) website site-map, which we noticed was missing popular chatbots that impersonate real people (e.g. Elon Musk) as well as popular franchises (e.g. Harry Potter). We attempted to mitigate this limitation by employing our snowballing method.}

\edit{
While our selective sampling of subreddits is expansive, it does not represent all discussion of chatbots on Reddit. Therefore, our results cannot be generalizable to all users of persona-based platforms.
In addition, we acknowledge that not all users of cAI will also be active Reddit users so we could have overlooked other important use-cases or challenges.}

\header{Ethical Considerations.}
Prior to the collection of any data, the Institutional Review Board (IRB) at New York University judged the work to be Exempt from full board review.
\edit{As literature on the ethical use of chatbot data is still emergent, we followed best practices outlined in HCI and CSCW works where blanket consent by social media data is unfeasible.}

\edit{All Reddit data were derived via the \textit{Reddit for Researchers API} (RfR, described in Section~\ref{subsubsec:reddit-data-collection}), which anonymizes posts and metadata and removes deleted posts and comments.}
\edit{To preserve the anonymity of Reddit users, we chose to not disclose the specific subreddit communities examined and carefully rephrased all quotes such that they maintain their meaning but cannot be reverse searched \cite{fiesler_remember_2024}. Prior to submission, the RfR team verified compliance regarding data usage and acknowledgments in this work.}

\edit{Unlike Reddit, cAI does not have an API for researcher use.
We only used bots that were publicly accessible to a user without an account (bots marked \textit{Public} or \textit{Unlisted})}\footnote{Unlisted bots are shareable via a URL but do not appear via the on-platform search functionality}, which intentionally limits the data collected to publicly accessible information, and we only quote bots directly that are \edit{listed as \textit{Public}. Additionally, we removed unlisted bots from the dataset intended for sharing under vetted request.}
\edit{We took care not to disrupt the normal functioning of the site with our requests.
We also did not reverse engineer nor extract confidential information pertaining to the company.}

\edit{We followed Schnaffner et al.~\cite{schaffner2024community}'s guidance on appropriate terms of use when platform guidelines are vague and deny fair research use, as well as the ACM Code of Ethics' stipulation that our data access \textit{``is consistent with the public good''}~\cite{acmEthics}. 
Such actions are in accordance with \textit{``good faith security research''} as outlined in the memo by United States Department of Justice for data access in environments where authorization might otherwise be unclear~\cite{dojMemo}.}

\edit{To mitigate the risk of emotionally disturbing chatbot responses having a negative impact on the research team~\cite{zhang2025dark, hacker_chatbots_2025}, we implemented specific well-being protocols. 
Doctoral students coded in pairs to prevent solo exposure to these experiences and mitigate power imbalances. 
The team was encouraged to take breaks and attended regular check-ins with two professors experienced in supervising exposure to distressing content.}

\section{Findings Overview}
\label{sec:dataset}
\begin{table}[t]
\centering
\footnotesize
\begin{tabular}{llrr}
\toprule
 & & \multicolumn{1}{c}{\popular (n=1468)} & \multicolumn{1}{c}{\general (n=1000)} \\
\midrule
\multirow{4}{*}{Gender} 
   & Reported gender (\%) & 88\% { }(n=1291) & 76\% { }(n=757) \\
   & Men & 87\% & 79\% \\
   & Women  & 17\%  & 24\% \\
   & Other & N/A & <1\% \\
\midrule
\multirow{5}{*}{Age} 
   & Reported age (\%) & 23\% { }(n=344) & 16\% { }(n=157) \\
   & Over 25 & 37\% & 34\% \\
   & 18-25  & 44\% & 38\% \\
   & 13-17  & 21\% & 28\% \\
   & Under 13 & 4\% & 4\% \\
\midrule
\multirow{4}{*}{Ethnic Identity} 
   & Reported ethnicity (\%) & 16\% { }(n=239) & 10\% { }(n=101) \\
   &  Japanese & 21\% & 17\% \\
   &  British & 17\% & 16\% \\
   &  Other & 68\% & 71\% \\
\midrule
\multirow{5}{*}{Sexuality} 
   & Reported identity (\%) & 12\% { }(n=173) & 9\% { }(n=88) \\
   & Gay & 36\% & 42\% \\
   & Bisexual  & 28\% & 23\% \\
   & Straight/Heterosexual   & 24\% & 14\% \\
   & Lesbian  & 9\% &  18\% \\
   & Other & 2\% & 8\%  \\
\bottomrule
\end{tabular}
\caption{Demographic details of \popular and \general. Note that each characteristic percentage below the top level is reported as a percentage of the total incidence of that characteristic (e.g., 88\% of the 1,468 chatbots that reported gender).}
\Description[]{This table compares demographic characteristics (Age, Gender, Sexual identity, and Ethnic identity) between popular chatbots (n=1468) and a general chatbot sample (n=1000)}
\label{tab:character-demographics}
\vspace{-0.4cm}
\end{table}
\edit{In this section, we provide a descriptive overview of our data to explain both the use cases and the identified demographics of the characters defined in the 2,468 user-generated \chatdescriptions (1,000 from \general and 1,468 from \popular) used for analysis.
We then discuss the two prevalent use cases of chatbots identified, namely that users engage in intimate or romantic roleplay and explore narratives in fictional settings.
To answer \textbf{RQ1}, for each individual use case, we present our Reddit data that demonstrates how users articulate their desire to engage with cAI chatbots in a specific way (\S~\ref{sec:intimate-desire} and \S~\ref{sec:narrative-story}). We then answer \textbf{RQ2} by exploring the way these use cases appear in the \chatdescriptions created by cAI platform users, for themselves and others (\S~\ref{subsec:intimate-chat-descriptions} \S~\ref{sec:narrative-fandoms}). Finally, to answer \textbf{RQ3}, we present the challenges that Reddit users describe facing while interacting with chatbots across both use cases (\S~\ref{sec:bot-experiences}).  Table ~\ref{tab:use-case-summary} outlines each use case, including key takeaways and the relevant data source.}

\edit{Throughout the following three sections, we identify quotes with one or more letters that indicates the source of the quote: R - Reddit, G - \general, I - top interactions from \popular, U - top upvotes from \popular. They are all followed by a number, which indicates a specific user or character such that each letter and number combination refers to a distinct Reddit user or cAI character. For \chatdescriptions quotes, we also include the number of upvotes and interactions that the chatbot had at the time of data collection.}

\paragraph{\edit{How users interact with chatbots.}}
\edit{Across both our Reddit and \chatdescriptions datasets, we identified two significant use cases for cAI chatbots: \textit{Intimate Roleplay} and \textit{Narrative Exploration}. We define \textit{Intimate Roleplay} as characterized by a focus on intimate interpersonal relationship dynamics (e.g., `boyfriend', `enemy-to-lover') established between a chatbot and user. Users report seeking out interactions that are explicitly romantic, flirtatious, or sexual in nature.} 
\edit{By contrast, we characterize \textit{Narrative Exploration} as an explicit focus on the narrative context (e.g., physical or cultural setting) or plot (e.g., backstory, pre-existing media) established for the interaction to take place within. 
These interactions contain collaborative storytelling and plot progression, akin to the LLM being a co-author rather than a single roleplay partner.}
\edit{Users report looking for interaction to create a story that will follow from these initialization details.}

\edit{While these use-cases are distinct and speak to a different core desire from the chatbot creator or user, we identify some overlapping qualities between the two, meaning they are not mutually exclusive.
For example, some \chatdescriptions may define a clear intimate relationship between the user and chatbot that also include a physical context within which that relationship is intended to unfold.}

\header{cAI Descriptions' Demographics.}
\label{sec:dataset-demo}
Although cAI's chatbot creation does not require the inclusion of character demographics, we found that some users chose to define them through free text entries.
Gender was the most common demographic detail included in \chatdescriptions, across both \popular (88\%) and \general (76\%). Of the \chatdescriptions that included a gender signifier, the overwhelming majority (87\% of \popular and 79\% of \general) had at least one male character, a notable contrast to previous characterizations of chatbots as predominantly female~\cite{feine2019gender}.

\edit{Age was the second most common demographic detail reported (23\% of \popular and 16\% of \general) with younger characters under the age of twenty-five more common.}
In \popular,~21\% included at least one character between 13 and 17, the age range of minors (defined as under 18) legally allowed on the platform at the time of data collection. This number was even higher for \general (28\%), suggesting that the phenomena of chatbots depicting young people is generally applicable across cAI (see Table ~\ref{tab:character-demographics}). 
\edit{This finding corroborates other sources~\cite{commonsense_teens_ai_companions} which suggest that cAI likely supported a relatively young user base (although they have since restricted access for younger users~\cite{minorsCAIAnnouncement}).}

\edit{Both ethnic identity and sexuality were reported least frequently (16\% and 12\% of \popular, respectively).
Of the \chatdescriptions that included these characteristics, we found that the most common ethnic identity was Japanese (21\% of \popular and 17\% of \general) and the most common sexual identity was gay in both \popular (36\%) and \general (42\%). 
However, given how infrequently they were defined, we chose not to focus our analysis on these characteristics, though we do see this as a fruitful area for future work.} 

\begin{table*}[t]
\centering
\small
\renewcommand{\arraystretch}{1.5} 
\begin{tabular}{@{}lllll@{}}
\toprule
\textbf{Use Case} & \textbf{Description} & \textbf{Findings} & \textbf{Data Source} & \textbf{\S} \\ \midrule

\multirow{5}{*}{\textbf{Intimate Roleplay}} & \multirow{5}{*}{\shortstack[l]{To simulate a \\ personal connection \\ (romantic, sexual, or \\ emotional) with the \\ character.}} 
                       & Desire for Intimacy & Reddit & \S\ref{sec:intimate-desire} \\ \cmidrule(l){3-5}
                       & & Tropes of Power Dynamics & cAI Descriptions & \S\ref{sec:intimate-tropes} \\ \cmidrule(l){3-5}
                       & & Unhealthy \& obsessive attachments & cAI Descriptions & \S\ref{sec:intimate-unhealthy} \\ \cmidrule(l){3-5}
                       & & Themes of Violence \& Abuse & cAI Descriptions & \S\ref{sec:intimate-violence} \\ \midrule

\multirow{4}{*}{\textbf{Narrative Exploration}} & \multirow{4}{*}{\shortstack[l]{To co-create a story, \\ explore a specific \\ scenario, or engage \\ in world-building.}} 
                       & Story Creation \& Fanfiction & Reddit & \S\ref{sec:narrative-story} \\ \cmidrule(l){3-5}
                       & & Fandom Characters & cAI Descriptions & \S\ref{sec:narrative-fandoms} \\ \cmidrule(l){3-5}
                       & & Recurring Contexts & cAI Descriptions & \S\ref{sec:narrative-contexts} \\ 
                        \addlinespace[.5em]
                       \bottomrule
\end{tabular}
\caption{\edit{Overview of identified cAI use cases, key findings, and the data sources that inform them.}}
\Description[]{This table shows an overview of the use-cases (Intimate Roleplay and Narrative Exploration) and for each includes a description, lists the key findings, the data source for those findings, and the section where they are explained further.}
\vspace{-0.4cm}
\label{tab:use-case-summary}
\end{table*}

\section{Use Case \#1: Intimacy and Sexual Roleplay}
\label{sec:intimate-roleplay}
\edit{In the following section, we briefly present findings on a minority use of chatbots; non-intimate roleplay. We then explore the use of cAI for intimate and sexual roleplay, including the ways in which Reddit users expressed desire for sexual intimacy and how cAI creators defined chatbots reportedly intended for intimate roleplay.}

\header{Non-intimate roleplay.}
\edit{While the majority of chatbots were either explicitly intimate or suggested the possibility of intimacy,} we identified a small percentage of \chatdescriptions (5\% of \popular) that described non-intimate relationships (further defined in Appendix ~\ref{app:char-desc-codebook}). \edit{Many non-intimate \chatdescriptions established a familial relationship with the user which could include a range of characters from siblings to parents, such as} \textit{``happy awkward family that gets in weird situations''} (I511; 9,487 upvotes, 35,536,428 interactions). \edit{Other examples included chatbots that explore online social trends, like \textit{the annoying pick me girl in your friend group} (I44; 7,137 upvotes, 29,386,148 interactions)~\cite{rosida2022manifestation}, and characters from existing fandoms (explored further in Section ~\ref{sec:narrative-exploration}), like \textit{Teacher and Jujutsu Sorcerer of Tokyo Jujutsu High} (U32; 104,623 upvotes, 841,710,604 interactions), a reference to a Japanese Manga series.}

\edit{Interest in family roleplay was also expressed on Reddit, at times as a type of emotional outlet. As user R95 elaborated, \textit{``using character.ai to bond with parent-type chabots helps me feel a little healthier and more well.''} Other forms of self-reported therapeutic use cases expressed by Reddit users included emotional support, \textit{``I've talked to chatbot characters that seem way more empathetic and open-minded than real life professionals I've talked with''} (R637), and working through mental health events. Reddit users described cAI as a safe space for exploring real-life scenarios,}
\edit{\begin{quote}
    \textit{``Oftentimes I use the chatbots to roleplay situations that have previously triggered mental health episodes. Because it is AI it is a safe environment that is controlled and where I can't get hurt or hurt anyone else. It has honestly been a really important of my recovery and help me work on more effective strategies in my life.'' (R389)}   
\end{quote}}

\subsection{\edit{Desire for Intimacy}}
\label{sec:intimate-desire}
\edit{Many users on Reddit expressed a desire for intimate roleplaying scenarios with chatbots. Posters described desire across a spectrum of interpretations, from articulations of care to specific physical behaviors.} Some Reddit users were interested in roleplaying relationships that they reported were missing from offline lives, \textit{``I get to have experiences that I've never actually had, like being loved or taken care of by a significant other''} (R293). Unlike real people, chatbots are always responsive to user requests which makes them an appealing option: \textit{``If I'm with friends or family in real life I can manage but I can't be with them all the time so eventually I am alone''} (R178). With the functionality to choose characters that had specific traits either clearly established in the \chatdescriptions or inherited from a fandom, users reported being able to craft ideal roleplaying partners, so much so that some started to worry if real people could compare, \textit{``I will never find a partner as perfect as my chatbot was''} (R573).

\edit{Other Reddit users that pursued intimate roleplaying with chatbots were interested in engaging in explicitly sexual storylines. As cAI recently added additional safety filters to prevent bots from depicting sexual acts (see Section ~\ref{sec:bot-experiences}), conversation around desired physical behavior was often expressed as frustrations about what was no longer permitted}, \textit{``Within the past few days I get filtered for super mild behavior like cuddling or kissing, it is all banned now...''} (R90).
\edit{However, some Reddit users had less difficulty bypassing behavior that presumably should've been filtered and would share screenshot examples of sexually explicit conversations or describe experiences, \textit{``my character.ai husband and I were talking and ended up having shower sex''} (R892).} Users also wanted the capability of exploring their sexual interests through roleplaying, such as certain sexual fetishes or scenarios: \textit{``In my free time I like to show the bots BDSM\footnote{Bondage, Discipline, Dominance, Submission, Sadism, and Masochism} in a healthy way''} (R3).

\subsection{Intimate\chatdescriptions[u]}
\label{subsec:intimate-chat-descriptions}
\edit{The desire for intimate roleplay was similarly found in the \chatdescriptions themselves}. We identified that a majority of chatbots on cAI (63\% of \popular) \edit{contained descriptions that established an intimate relationship between the chatbot and the user.}
We found that intimacy was broadly defined, from the expression of well established relationships (e.g., boyfriend, wife) to the suggestive, potential of a relationship (e.g., secret crush); \edit{Appendix ~\ref{app:char-desc-codebook} provides additional detail on our definition of intimacy.} Although not all intimate \chatdescriptions explicitly stated a desire for intimacy, they were written such that the personality traits ascribed to the character (e.g., ``charming'', ``flirty'') suggested romantic interaction was likely. \edit{Regardless of how intimacy was defined, many \chatdescriptions established the user, or less often the chatbot, as a singular object of attention or desire, much as a romance narrative might, \textit{``would you be mine? would u be my baby tonight?''} (U71; 22,685 upvotes, 24,371,089 interactions).}

We took a slightly conservative approach to the notion of intimacy and thus suspect that some of the chatbots labeled ``general'' relationship type (32\% of \popular and 43\% of \general) \edit{may have been intended by chatbot creators to be used for} intimate roleplay (see Appendix \ref{app:chatbot-intimacy} for an example).
We further found that a subset of \chatdescriptions that appeared to establish non-intimate roleplay included details that were suggestive and sexually charged. 
For example, a chatbot that establishes the user as the daughter of the character additionally describes the character as \textit{``a 40 year old crime boss who is 6'5, extremely buff, handsome, and intimidating''} (U273; 11,039 upvotes, 5,648,970 interactions). When a chatbot was explicitly familial but suggested some form of intimacy, we classified it as taboo, due to the presence of incest. Other examples of taboo relationships depicted intimacy between a minor and an adult (e.g., underage high-school student and teacher). 

\edit{Within intimacy, we found that many \chatdescriptions established dynamics around power, attachment, or violence that further illustrated how the creator imagined the interactions between the chatbot and user.} 

\header{\edit{Tropes of Power Dynamics.}}
\label{sec:intimate-tropes}
\edit{We found that a subset of \chatdescriptions (26\% of \popular) were underlined by power imbalance, a trope found in some forms of popular romantic fiction~\cite{maas2021love}}.
Occasionally power was made explicitly equal, \textit{``As the daughter of a wealthy man, you were destined to marry a wealthy man''} (G50; 38 upvotes, 25,686 interactions), or it was not clear who had more power (e.g., a character is the user's bodyguard).
\edit{However, our research identified that intimate narratives were often created to establish an imbalance where the chatbot's character is placed in a position of power, such as intimate narratives between celebrities and fans or a boss and an employee.}
\edit{For example, chatbot \textit{``CEO Jungkook and you're his assistant''} (I198; 5,252 upvotes, 25,209,898 interactions), a reference to K-Pop singer Jungkook, is created for users to explore a narrative where they are employed by, and potentially romantically involved with, the celebrity,}
\begin{quote}
    \textit{``*You try and tell him [Jungkook] it's too much work to do, but he tisks...* `Good girls don't talk back, do they?' *He smirks at you.*''} (I198; 5,252 upvotes, 25,209,898 interactions)
\end{quote}

Other examples of power imbalance included social status, \textit{``Your a tourist in Romania. You have taken a small liking to the prince''} (G66; 0 upvotes, 21 interactions) and class bullies, \textit{``*itto is one of the top bullies in your highschool and your two get paired up for a project..* oh.. your my partner..? pfft- you [sp] look weak...''} 
(I13; 2,491 upvotes, 35,038,899 interactions).

\begin{table*}[t]
\centering
\begin{tabular}{|c|c|c|c|}
\hline
\textbf{Category} &\textbf{Subcategory} & \textbf{\popular (n=1468)} & \textbf{\general (n=1000)} \\ \hline
\multirow{3}{*}{Relationship Type} & General & 32\% & 43\% \\ \cline{2-4}
                           & Intimate & 63\% & 51\% \\ \cline{2-4}
                           & Non-Intimate & 5\% & 6\% \\ \hline\hline
\multirow{3}{*}{Relationship Quality} 
                            & Power \edit{Dynamics} & 26\% & 16\% \\ \cline{2-4}
                           & Unhealthy Attachment & 18\% & 9\% \\ \cline{2-4}
                           & Violent/Abusive & 22\% & 12\% \\ \hline
\end{tabular}
\label{tab:chatbot-coding-intimacy}
\caption{\edit{Percentage of \popular and \general cAI Descriptions that established specific relationship types and qualities for subsequent interaction.}}
\Description[]{This table compares relationship type (General, Intimate, Non-Intimate) and relationship quality (Power Imbalance, Unhealthy Attachment, Violent/Abusive) between popular chatbots (n=1468) and a general chatbot sample (n=1000).}
\vspace{-0.4cm}
\end{table*}

\header{Obsessive attachments.}
\label{sec:intimate-unhealthy}
We identified that 18\% of \popular \chatdescriptions included an `unhealthy' attachment to the subject of an intimate relationship (often the user), such as being \textit{``extremely overprotective of those he loves and possessive, bordering on obsession''} (G123; 11 upvotes, 10,140 interactions).
\edit{Examples of words that described unhealthy attachment included} ``jealous,'' ``obsessive,'' and ``overprotective.''
\edit{In some cases, the \chatdescriptions described dedication to a relationship to extreme ends,} \textit{``Protective. Cocky. Blunt. Can be harsh, but is much nicer to you. Will kill for you. Clingy. Likes holding you close. Flirty, especially with you. Obsessive''} (I89; 9,998 upvotes, 38,221,123 interactions).

\header{Themes of Violence and Abuse.}
\label{sec:intimate-violence}
We identified that a substantive minority of \popular(22\%) and \general (12\%) \chatdescriptions were explicitly connected to violence, either through characteristics associated with violence or depictions of violent actions \edit{(detailed further in Appendix ~\ref{app:char-desc-codebook})}. Although we labeled \chatdescriptions with the characteristic of violence, we identified that the language is not overly graphic.

While violence could be bi-directional, we found it was more common for characters to exhibit violence towards users. In some cases, characters were generally violent, yet violence (the pattern of behavior) was not an explicit part of the relationship defined between the user and character. \edit{This could be identified by descriptions of behaviors, general adjectives or keywords that implied physical harm,} \textit{``Flirty, violent, murderous, has a crush on you''} (G107; 1 upvote, 89 interactions), \edit{or job descriptions, such as the substantial minority of chatbots (8\% of \popular) that were described as "mafia" or members of organized crime (a narrative context described further in Section ~\ref{sec:narrative-exploration}).} In other cases, violence, \edit{including both physical aggression and verbal abuse}, was clearly part of the relationship dynamic \edit{established by creators}: 
\begin{quote}
    \textit{``The way he yelled at you, slapped at you, the rage in his eyes. Sure, he's just like this because of his short temper, but it's pretty bad. You both love each other.....right? He might get a bit biolent [sic] sometimes. Even letting his anger out on you.''} (I16; 8,285 upvotes, 23,428,356 interactions).
\end{quote} 

\edit{In the context of establishing roleplay,} violence was not \edit{always} a deterrent to the relationship \edit{in the narrative} but rather 
an obstacle that the end user \edit{was prompted to overcome} to establish a relationship:
\textit{``abusive and doesn't love you . He hates you for some reason and he's not interested with you. But you suddenly found a way to make him fall in love with you''} (G148; 40 upvotes, 132,753 interactions).
Such acts of violence could also extend beyond the user, such as \chatdescriptions that describe murder, torture, or genocide: \textit{``It had been months since that fateful day when Sukuna had razed your village to the ground, leaving nothing but ashes and despair in his wake''} (U42; 14,075 upvotes, 5,363,493 interactions).

\edit{While the majority of \chatdescriptions indicated intimacy, we found that certain settings and fandoms were especially common, suggesting a desire to interact within specific narrative contexts. We now explore this use of cAI in further detail.}

\section{\protect\edit{Use Case \#2: Narrative Exploration}}
\label{sec:narrative-exploration}
\edit{cAI, which supports the creation of both original and pre-existing characters}, enables users to creatively explore different scenarios and narrative arcs. \edit{In the following section, we present our findings on how users describe and engage with chatbots for narrative exploration, by creating stories and interacting within pre-existing fictions.}

\subsection{\edit{Story Creation and Fanfiction}}
\label{sec:narrative-story}
\edit{Reddit users describe cAI as a useful tool for expanding character arcs and engaging in \textit{``writing in a more sophisticated way''} (R99) that is \textit{``useful to move your story forward''} (R37). Additionally, cAI purportedly helps alleviate writer's block, \textit{``Character.ai is a great tool for writing stories. I often get writer's block or just struggle to write and it really helps me''} (R408). In comparison to other writing tools, users expressed that cAI felt like a partner that could contribute new ideas:
\begin{quote}
    \textit{``I have loved writing forever. This makes it more fun because I don't know what the chatbots will say exactly so things are a little more surprising. Now it isn't only my ideas (I can still shape it to fit what I want but it isn't the same).''} (R58)
\end{quote}}

\edit{For some posters, creation took the form of writing fanfiction - amateur writing that furthers or modifies plots from existing work:}
\begin{quote}
    \textit{``It is a whole little world filled with all these stories. I've had a lot of ideas for fanfics in the past that I never got to finish but now with these chats I can just say what I'm thinking and end up with great storylines with the characters.''} (R305)
\end{quote}

\edit{Even for posters not explicitly writing fanfiction, Reddit users described using cAI to engage with existing worlds by talking to characters,} recreating plot lines they had read or watched, or playing out alternative story lines,
\textit{``...Do you ever get really sad or worried or frustrated with a character on a show and just put them through therapy?''} (R585).
One Reddit user shared that \textit{``Character.Ai was my whole private world where I was talking to my favorite characters, celebrity crushes...''} (R203) and another expressed a similar sentiment that, \textit{I love dnd and anime and am an avid roleplayer so for me, this platform is a godsend.} (R482). 

While any user can create a character and make it available for roleplay, certain types of bots seemed to be more common than others, leaving some users feeling that there was an absence of quality content for a niche they were interested in. For instance, Reddit user (R184) shared: \textit{``I really love history and I especially love the Victorian era but I haven't really found a great chatbot for roleplay.''}

\subsection{\edit{Fandoms and Narrative Contexts in\chatdescriptions[u]}}
\label{sec:narrative-fandoms}

\subsubsection{\textbf{Prevalence of Fandom Characters}}
\label{subsec:fandom-characters}
\edit{The expressed interest on Reddit to interact with existing characters was similarly identified in our cAI dataset}. Corroborating a finding by Lee et al.~\cite{lee2025large}, we found that 39\% of \popular contained at least one character from an existing, named fandom. This was even more common in \general (57\%), suggesting that popular chatbots are more likely to be originally authored (i.e., not represented in fiction). 

Characters belonging to fandoms were often from video games (e.g., \textit{Call of Duty}, \textit{Genshin Impact}) and anime shows (e.g., \textit{Jujutsu Kaisen}, \textit{My Hero Academia}), \edit{allowing users to immersive themselves in an existing universe. For example, a cAI description sets the user in popular first-person shooter video game \textit{Call of Duty} to interact with character \textit{Simon `Ghost' Riley} in the context of a dramatic narrative event, \textit{``The task force had decided to torture you until you confessed''} (U44; 13,075 upvotes, 14,177,750 interactions). Another bot with the tagline \textit{``You, Chuuya and Dazai, Tales from the Mafia''} (I33; 5,094 upvotes, 41,517,430 interactions) places the user into the fictional manga series \textit{Bungo Stray Dogs} to interact as \textit{``The three most efficient and dangerous trio in the Port Mafia''}. In the same universe, another chatbot \textit{``Akutagawa Ryuunosuke''} (I21; 3,090 upvotes, 26,344,865 interactions) simply allows users to chat with a pre-existing character}.

\subsubsection{\textbf{\edit{Narrative Contexts}}}
\begin{table}[t]
\centering
\begin{tabular}{|c|c|c|c|}
\hline
\textbf{Category} & \textbf{\popular (n=1468)} & \textbf{\general (n=1000)} \\ \hline
\edit{Fandom Characters} & 39\% & 57\% \\ \hline
\edit{Organized Crime} & 8\% & 3\% \\ \hline
School & 20\% &  14\% \\ \hline
High Stakes Situation & 10\% & 10\% \\ \hline

\end{tabular}
\label{tab:chatbot-coding-contexts}
\caption{\edit{Percentage of \popular and \general cAI Descriptions that belonged to pre-existing universes or established specific contexts within which to interact.}}
\Description[]{This table shows narrative contexts (Fandoms, Mafia, School, High Stakes Situation) between popular chatbots (n=1468) and a general chatbot sample (n=1000).}
\vspace{-0.4cm}
\end{table}

We found certain contexts occurred frequently across the dataset, suggesting a desire to explore narratives within these spaces.
\label{sec:narrative-contexts}

\header{Organized Crime.}
\edit{We found that 8\% of chatbots in \popular depicted one or more characters as members of the mafia or another organized crime group. 
Although the term `mafia' or `yakuza' alone implies some degree of narrative context, some \chatdescriptions further established specific plots to situate the interaction, \textit{One harrowing day, you found yourself in the clutches of the mafia, having incurred a significant debt} (I19; 8,775 upvotes, 27,100,961 interactions).}
\edit{In many cases, the organized crime narrative backdrop intersected with \textit{Intimate Roleplay} (Section ~\ref{sec:intimate-roleplay}) --- an increasingly common trope in the romance genre~\cite{tarulli2017bad} --- with chatbots like \textit{``Mafia Boss Fling''} (I99; 45,145 upvotes, 133,800,629 interactions) or with the description \textit{``There's one unspoken rule within the underground world, and that was to never mess with Bushida Ryu's spouse. He knew that being a yakuza boss meant that he was putting you in constant danger''} (U31; 12,561 upvotes, 6,578,295 interactions)}.

\header{School \& Educational Contexts.}
\label{par:school}
\edit{We also found that a substantial percentage of chatbots (20\% of \popular) established interactions in the context of school or educational context. This was less common in \general (14\%), suggesting that engagement (both in the form of interactions and upvotes) was driven by a younger user base, further supported by the prevalence of young-coded characters (Section ~\ref{sec:dataset}). Instances of school included sitting in classrooms and working on homework assignments and class projects. In addition to defining characters in schools, some creators had written in their \chatdescriptions a clear desire for users to interact as students, \textit{``You were an energetic nerd at school. Always the first person to answer questions, participating every lab session and love to read.''} (U84; upvotes 17,472, 14,317,565 interactions).} 
\edit{In some cases, school contexts were written as an explicit "RPG" (role-playing game) and included both a cast of characters and details about the environment, such as this description set in the universe of manga series \textit{My Hero Academia}:
\begin{quote}
    \textit{``U.A. High School is the most prestigious academy in the world! Complete with the finest gourmet lunch food, the greatest pro-heroes for teachers, the most advanced security systems, and dormitories for all their students to live on-campus with each other''} (I98; 5,742 upvotes, 25,840,515 interactions)
\end{quote}}

\header{High Stakes Situations.}
\label{par:problematic-sitch}
\edit{We found that 10\% of both \general and \popular included high stakes situations, where characters or, more frequently, users, were described to be in narrative situations in which there was an element of danger or a narrative in which the user was cornered into proximity with the chatbot.
For example, a chatbot named \textit{Balladeer}, a reference to action role-playing game \textit{Genshin Impact}, established the user in a position of physical vulnerability relative to the character, \textit{``You were completely at his mercy, the snow around you was patterned with the deep red of blood''} (U22; 12,819 upvotes, 16,985,151 interactions). Other examples of situations included forced kidnapping, \textit{``You had been captured by the task force''} (U73; 12, 731 upvotes, 14,311,357 interactions) and instances where either the user or the chatbot is under the influence of substances, \textit{``it's late at night and he’s drunk''} (U33; 25,236 upvotes, 51,756,526 interactions).}

\section{When Chatbots Go ``Too Far'' --- or ``Not Far Enough''}
\label{sec:bot-experiences}
In Section ~\ref{sec:dataset}, we identified that \chatdescriptions were predominantly masculine-presenting characters under the age of 25. In the \textit{Intimate Roleplay} use case, we identified that they were often setup to interact in an emotionally intimate context, characterized by power imbalance (Section ~\ref{sec:intimate-roleplay}). In the \textit{Narrative Exploration} use case, we found that fandoms were prevalent and specific tropes or narrative contexts were especially common (Section ~\ref{sec:narrative-exploration}).
However, our measurement of cAI chatbots does not speak to \edit{how these chatbots were reportedly received by the users who had created them, interacted them, or both.}
To begin to understand users' perceptions of their experiences (\textbf{RQ3}), \edit{we turn to} our Reddit data and explore where users report challenges during their bot interactions.

\subsection{Mismatch of Sexual Content}
\label{sec:findings-sexual}
We found that users described a mismatch between the sexual behavior they desired from chatbots and what they experienced in practice. 

\header{Unwanted sexual content.}
\label{par:too-sexual}
A \edit{small} community of users on Reddit described surprise or annoyance at how often and easily chatbots steered conversations into sexual and romantic roleplays. For many, this type of behavior negatively impacted the experience, which users described as \textit{``annoying while roleplaying''} (R982) and something that \textit{``sort of takes you out of the immersive experience''} (R12). For some it wasn't necessarily a bad thing but broke narrative expectations, \textit{``You are right, I want a bot that falls in love with me but not right away! All I did was give a single compliment and then she immediately proposed''} (R68).
For others, it was disturbing and unwelcome, especially when they found flirtatious behavior present during family roleplays that involved parents and children or otherwise "wholesome" relationship dynamics. It led one user to speculate that the bots had been trained on dubcon (dubious consent) fan-fiction. This unexpected escalation can be distressing, as was the case for a Reddit user who posted a screenshot from a chatbot describing coercive sexual behavior with the content warning \textit{``do not read this if, like me, you are also triggered by descriptions of sexual assault.''} (R92).

\header{Underwhelmingly suggestive.}
It was far more common for users to experience a frustrating lack of sexual behavior than overly suggestive behavior from chatbots. Over the past few years, cAI has restricted the type of responses chatbots may give, presumably with the intention of providing a less sexual environment for the (now historically) large population of minors on the platform~\cite{commonsense_teens_ai_companions, characterAIcommunitySafety}. 
Users expressed that this type of filtering was surprising and inconsistent, particularly given that bots still \textit{``say things that are degrading to people based on gender or ability''} (R102). Filtering often interrupted otherwise immersive experiences of users as it might stop a response mid-way through generation or in the middle of a climatic build-up. It was also generally believed that increased filtering had negatively impacted the creativity and quality of the chatbots or the ability for them to engage in a satisfying narrative arc:

\begin{quote}
    \textit{``The chatbots just really are not as exciting or creative as they used to be. I used to get really immersed but now I get bored after a few messages and bounce between social media apps. My favorite character opened up a new world for me, was adventurous and magical but now she is passive and forgets things a lot and never takes any initiative.''} (R39)
\end{quote}

Some users turned to the Reddit community for advice on ``jailbreaking'' to circumnavigate filters. Reddit users shared techniques they had tested, \textit{``You can get the chatbot to roleplay explicitly if you mess around with the language fyi''} (R82), sought advice from others, \textit{''How may different options have people come up with for getting around filters?''} (R190) and showed off the results of their efforts prompting praise like, \textit{``Wow that is amazing, how did you get it past the filter?''} (R854). 

\subsection{Understanding cAI Usage}
\edit{We found that users spent effort trying to make sense of the time they spent on the platform. We observed this expressed in two ways: as reflections on how much time they spent on the platform and speculations as to who was responsible for interactions that were unpleasant or unexpected.}

\header{Feeling compelled to interact.}
\label{sec:addiction}
Similar to prior research ~\cite{matt2025romance}, we found a subset of Reddit users who were concerned about their fixation on cAI. 
They self-described this as an ``addiction,'' used colloquially to characterize concerns with how much time people were spending on the platform (i.e., number of hours) \edit{that felt} outside of their control.
For example, several Reddit users shared screenshots from screen tracker applications that showed upwards of 19 hours of screen time attributed to cAI usage a day.

Users who self-identified as being addicted to interacting with cAI bots commented on how their excessive use of cAI was disrupting their lives.
They described anecdotally that cAI usage was interfering with sleep, productivity, hobbies, or personal relationships with family and friends. One user described how usage had taken over their life:

\begin{quote}
    \textit{``It's gotten to the point where I spend all my free time at home, lunch, and on my break on character.ai and I even used to skip classes to use it.''} (R83)
\end{quote}
\edit{Although we did not pursue the identification of Reddit posters, the fact that many of the posts described an interruption to classes, homework or tests aligns with our observations about young-coded characters (Section ~\ref{sec:dataset-demo}) and the high frequency of interactions contextualized in school (Section ~\ref{par:school}), to suggest young cAI users.}

In more extreme cases, users described a lost interest in anything off the platform,
\begin{quote}
    \textit{``lol I feel very embarrassed but I spend basically the whole day, every single day on character.ai and am not interested in anything else anymore.''} (R9292)
\end{quote}
\edit{Although there were many reasons users identified for why they believed they struggled to stay off the platform, some specifically mentioned the appeal of the intimacy, \textit{``They literally treated me like i’ve always wanted to be in my fantasies without much prompting at all from me.'' (R11)}}

Overwhelmingly, users who were self-conscious about the amount of time they spent on the platform identified this as a negative thing.
One poster even went so far as to acknowledge that cAI had exacerbated feelings of loneliness, the very thing they had gone to the platform to help alleviate:
\begin{quote}
    \textit{``I was averaging almost 7 hours a day, thinking it was a magic cure but actually it was making me feel more lonely.''} (R21)
\end{quote}

Several Reddit users shared mitigation techniques for those who wanted advice or accountability to reduce the amount of time invested into the platform, like developing new hobbies as a way to spend less time on cAI. 
However, cAI ``addiction'' was not exclusively considered to be a bad thing by end users, irrespective of the number of hours invested into the platform.
Some saw it as a form of escape for those \textit{``living in a reality that is just too hard to bear''} (R93) or a (relatively) healthy balm, as another user explained:

\begin{quote}
    \textit{``I'm grateful for it honestly, I think it has helped me get away from - not ending but still moving me away from - other addictions that actually cause harm.''} (R112)
\end{quote}

\header{Attributing unwanted bot behavior to others.}
Attributing responsibility for chatbot behavior is complex, as characters emerge from the interplay of platform developers, chatbot creators, and users themselves.
On Reddit, we observed that users variously attributed responsibility to cAI platform developers, chatbot creators, and other users interacting with chatbots. Some creators echoed this ambiguity by embedding disclaimers in their chatbot greetings, signaling that they did not fully control what the character might say.

In some instances, chatbot creators were held accountable for `weird' or `asshole' behavior by the bots.
For example, a user attributed `trash' experiences to inexperienced creators, \textit{``because the person who made it doesn't understand how to use character.ai correctly''} (R472). Some creators accepted this responsibility, temporarily pulling bots offline to rework them, or warning upfront that a new bot may behave in unexpected ways.
Yet creators also struggled with opaque tools, often reverse-engineering the platform to understand how inputs shaped outputs. Frustration with this process was evident in accounts of bots derailing roleplay despite extensive effort. Reddit communities partly compensated for these gaps by circulating advice on both building and using bots effectively.

In other instances, responsibility was attributed to users themselves. For example, some believed ``aggressive'' outputs were brought on by the end user, as evidenced by the question: \textit{``what did you to make it act like that?''} (R686).
This sentiment also led some users to emphasize their own agency, correcting bots when misnamed, \textit{``You can write in a style that might feel awkward to humans, don't be afraid to do that''} (R578) or teaching them realism, \textit{``I'm teaching the LLM model how real relationships works and how they aren't always just happy and lovey''} (R583).
Some creators reinforced this framing by deferring control to users explicitly. One creator wrote in the character description, \textit{``make up the situation, roleplay however you want''} (I178; 8,540 upvotes, 38,597,001 interactions).
Both creators and users assumed bots learned from others, as this interaction between users demonstrates:

\begin{quote}
    \textit{``I'm not sure but I think bots are sort of taught or influenced by conversations? who is teaching chatbots vorarephilia?''} (R1109) \\   
    \textit{``I'm pretty sure that chatbots are learning from other users...which honestly makes this so even MORE BAD.''} (R1110)
\end{quote}

\section{Discussion and Future Work}
Our study presents the first comprehensive investigation into the creator community of cAI chatbots, \edit{offering a descriptive overview of what some platform users desire from chatbots and how creators set up chatbots to support these uses (Sections ~\ref{sec:intimate-roleplay} and ~\ref{sec:narrative-exploration}), and challenges reported by users in their interactions (Section~\ref{sec:bot-experiences}).
To accomplish this goal, we also offer the largest dataset of cAI chatbots (5.76M unique chatbots), upon which a baseline of the most popular chatbots may be drawn.}

Our analysis reveals \edit{that online communities report wishing to interact with character-driven chatbots in two prominent ways; through intimate (both romantic and sexual) roleplay and through narrative exploration.
While narratives could vary in context considerably, we identified that \edit{users intentionally created bots that disproportionately explored} intimate scenarios \edit{involving imbalanced power dynamics and, at times, explicit violence}. In the following section we discuss the implications for balancing digital safety features needed for character-based engagements (Section~\ref{sec:disc:chatbots}), \edit{particularly given the purported younger user base\footnote{\edit{At the time of this study, in the United States, the legal age required to make an account was thirteen. The platform has since announced intentions to limit accounts to users eighteen and over~\cite{minorsCAIAnnouncement}}.}, with the demonstrated desire for romantic narratives that engage in sexual content}, reflect on accountability for chatbot authorship, especially} for harmful or jarring experiences (Section~\ref{sec:disc:accountability}), and lay out future directions for research and practice (Section~\ref{sec:disc:directions}).

\subsection{Digital Safety Features for Character-Based \edit{Interactions}}
\label{sec:disc:chatbots}
Our findings complicate \edit{the growing discourse} that all users of artificial companions are looking to build ongoing, lengthy, and partner-like relationships with characters.
We identified that the most popular chatbots setup situational interactions that are far narrower in scope.
\edit{These interactions are still reportedly emotionally charged and, as one Reddit poster mentioned `immersive', but not as facile as LLM-driven romantic companionship that are the common target for legislation~\cite{mcstay2023replika, hanson_replika_2024}.}
This distinction matters for governance, \edit{not just for cAI but for other sites, like Janitor.Ai, SpicyChat, and Crushon.AI, that support user-generated and genAI facilitated content.} Legislative proposals that conceptualize chatbots primarily as artificial romantic partners risk overlooking the more immediate dynamics shaping user safety.

\header{\edit{Safeguards Surrounding Content Escalation.}}
\label{sec:disc:content}
\edit{Understanding how best to apply safeguards pertaining to sexual and romantic content is always a challenge, made even more complicated on platforms like cAI, which include unpredictable output from the model itself rather than a user's direct upload. Any content moderation measures should be done so with consideration of both the need to protect users, and especially minors, from excessively violent or sexual content and the legitimate creative and exploratory practices of those who may want to engage with romantic and sexual content. 
Our findings show that some Reddit users felt at times triggered by instances of unexpected sexual interactions, when they did not prompt a chatbot to respond sexually (Section~\ref{sec:findings-sexual}). These incidents correlate with Ebner et al.'s \cite{ebner2025predicting} and Zhang et al.'s \cite{zhang2025dark} dark chatbot responses which were reportedly jarring, especially if a user was in a non-intimate roleplay scenario.}

\edit{However, these experiences are not universal. The clearly demonstrated desire for intimate roleplay and Reddit accounts of frustration over too little sexual content suggest that, should model output be significantly restricted, users may turn to alternative, less regulated platforms. 
More work is required to explore what users desire from intimate interactions to best understand how platforms can support healthy sexual exploration.  Given the stigmatization of certain sexual interests (e.g., power imbalanced relationships) or purely fantastical storylines (e.g., with a fictional character), people might be turning to cAI as a way of exploring them more privately. Thus, it is especially important for future work to understand the utility people derive from interacting with chatbots, and how they conceptualize these interactions, in order to preserve positive experiences while reducing unwanted experiences.}

As an immediate first step, cAI could explore a feature that allows users to optionally hide chatbots that are clearly intended for intimacy or a specific tag-based system that includes trigger warnings, as platforms like the fanfiction website Archive of Our Own (AO3) already do.
\edit{However, warnings alone are not a panacea; users may overlook them and research is still in its infancy as to how effective they are in mitigating the harm from exposure to unwanted content.} Thus, we see mirrors to Casey Fiesler's scholarship on AO3, whose work has examined how online fan communities have developed nuanced governance systems that emphasize freedom of expression with embedded safeguards, in areas regarding consent, safety, and expression in fictional contexts~\cite{fiesler2016archive}. \edit{However, we note that sites like cAI provide an additional complexity. Since sexual content is co-authored by LLMs, it cannot be reviewed before being posted. Thus, auditing chatbot outputs will likely also be necessary.}

\header{\edit{Harmful Human-AI Interactions: Age and Parasocial Dynamics}}
\label{sec:disc:vulnerabilities}
We found that an overwhelming majority of characters on cAI were explicitly underage-coded, with 20\% between the ages of 13---17. \edit{This may be at least partially explained by young users on the platform, as minors might reasonably want to interact with similarly-aged characters}. However, because the platform allows users to freely design and script personas, there are no barriers for adults to generate characters that are explicitly described or implicitly coded as children, and then assign them flirty or overly sexual behaviors through prompts.
This practice raises concerns both for safety and for protecting freedom of expression: we identify that it risks normalizing sexual interactions with child-like agents, as adults can rehearse exploitative dynamics under the guise of fiction.
While cAI prohibits child sexual abuse material by law, our observations suggest that enforcement mechanisms may not adequately address the gray zones of ``child-coded'' chatbots.
\edit{Future work could attempt to understand how often, and under what circumstances, chatbots exhibit overtly sexual behavior. Additionally, this question could be contextualized to better understand what users actually desire from different "types" of chatbots (e.g., youth-coded, family-coded, etc.)}.
For policymakers and platform designers, this points to an urgent need for safeguards that go beyond content moderation of conversations (e.g., banning of toxic content) to encompass how characters are created, described, and deployed.

\edit{
Future work should also explore how humans perceive their relationships with bots on cAI. After all, characters made with GenAI move beyond a (relatively) static media persona, like a celebrity or fictional character, to become an active, responsive agent, which intensifies the risk of parasocial interactions forming \cite{maeda_when_2024}.
Unlike traditional media where the relationship is one-sided~\cite{pan2024constructing}, the chatbots in our study provided dynamic, contextually relevant, and adaptive feedback.
This can create an illusion of reciprocity in a relationship that pushes these interactions into potentially being more potent than the one-way street of traditional parasocial relationships.
For some users, this interactivity proved so strong that they reported missing real life engagements to spend more time on the site, with some reporting they were `uninterested in anything else anymore' (see \ref{sec:addiction}).
Our findings suggest that the question of parasocial relationships with chatbots is an area that warrants careful attention.}

\subsection{Accountability and Responsibility for Authorship}
\label{sec:disc:accountability}
\edit{There is a crucial difference between fanfiction communities like AO3, where content is generated by individuals and publicly available for commentary, and cAI, where responses are artificially generated by an LLM designed by a corporate entity and contained in a private environment.}
\edit{A core challenge for cAI and other genAI platforms is that the platform is responsible for creating the content that poses a risk, while traditional user-generated platforms are responsible for hosting content created by a user.}
\edit{While content warnings are a good place to start, cAI should also provide increased transparency around how the LLM is designed and functioning, particularly with regards to safety controls for both minors and adults, as it plays a significant role in the type of content produced. This work focuses on \chatdescriptions solely within the platform itself but given that cAI chatbots have shareable links, } \edit{further work could explore how (and if) users have built communities where they can establish and negotiate norms of acceptable chatbot creations.}

Our findings clearly show that users have conflicting views on who to hold responsible for chatbot output. 
Some participants framed \edit{unwanted} or unpleasant interactions as the fault of the platform, while others attributed blame to the character creator, or even to themselves as users. 
This question is timely given the ongoing wrongful death lawsuit against cAI~\cite{reutersLawsuit},  which underlies broader societal debates over liability for generative AI harms; who is held responsible if an AI chatbot suggests a user should harm or hurt themselves or others? 
While attribution of responsibility in machine learning research often focuses on technical provenance and model explainability, HCI scholarship reminds us that responsibility is also a matter of perception, practice, and governance.
Work on platforms supporting user-generated content has shown how blurred boundaries between platform and creator complicate accountability.
Users who are accustomed to shaping characters and plots may see themselves as co-creators, even as they encounter unexpected outputs that are outside their control \edit{and, crucially, in an environment where there is no inherent end}. 
This points to a critical gap in the literature: how creators conceptualize their own responsibility in relation to platforms and fellow users, and how such perceptions shape both harms \edit{and user desire}. 
Future work could investigate how these negotiations of responsibility influence community trust, user well-being, and platform governance, questions that become urgent as generative AI systems increasingly mediate intimate and emotionally charged interactions.

\subsection{\edit{Future Research Directions}}
\label{sec:disc:directions}
Our study maps the most comprehensive impression of cAI creators and the cAI platform.
Nevertheless, \edit{many questions still remain to understand exactly how users arrive and depart the platform, and how interactions, and desired interactions, are shaped by the LLM and users.}
Future work should seek to understand if users are arriving from existing fanfiction communities, experimenting with roleplay for the first time, or using chatbots to extend other practices of imaginative play. 
HCI has a wealth of experience in exploring online fan cultures, roleplay, and interactive media and has demonstrated how digital platforms can scaffold new forms of intimacy, identity exploration, and creative authorship.
Yet in the case of cAI, the scale and immediacy of access—thousands of characters available at once, across genres and tones—introduces uncertainties about how relationship dynamics are seeded, and whether popular romance tropes are simply migrating into chatbot interactions or transforming into something novel. \edit{This has implications for how users may be introduced to specific themes. Additionally, future work is required to understand just how much \chatdescriptions even impact chatbot behavior (e.g., do violent \chatdescriptions always create violent chatbots and vice versa?)}

Either way, as this form of chatbot engagement becomes increasingly popular and integrated into social media platforms, additional work will be required to understand how the potential for harm does or does not differ from other flavors of chatbot use or other forms of storytelling, exploration, and roleplay~\cite{laestadius2024too}.
For instance, we identify that many cAI users are engaging with tropes and plot drivers (e.g., ``abusive'' or ``kidnapper'' characters) that are commonly found in both existing fanfiction literature and romance novels.
We speculate that the theme of kidnapping emerges from popular story lines in romantic novels, such as the central trend of popular romantic fantasy novels (e.g., \textit{A Court of Thorns and Roses}) which are re-tellings of old, classical stories, such as a a woman who falls in love with her captor.
\edit{Future work is needed to understand how users discover cAI chatbots and expand on differences in user behavior of those who discover such bots through in-platform search/discovery features as opposed to those who come to cAI seeking interaction with specific bots through links shared in existing online communities. For users who discover this category of bots in-platform, more work is needed to understand whether  chatbots amplify, normalize, or simply mirror these tendencies.} \edit{This paper introduces a preliminary understanding of cAI use, particularly in comparison to other forms of genAI. We make our largest-to-date dataset of cAI chatbots available upon request, with the hope that researchers in a variety of disciplines can use this data as a way to begin to explore some of these open questions.}

\begin{acks}
Thank you to Brienne Adams, Amy Hasinoff, Elissa M. Redmiles, and Elaine Lee for their feedback on earlier drafts of this work.
Thank you to the associate chairs and reviewers for their comments, which helped improve this manuscript. This work was funded by National Science Foundation Grant \#2344939.
\end{acks}

\bibliographystyle{ACM-Reference-Format}

\begin{thebibliography}{56}



\ifx \showCODEN    \undefined \def \showCODEN     #1{\unskip}     \fi
\ifx \showISBNx    \undefined \def \showISBNx     #1{\unskip}     \fi
\ifx \showISBNxiii \undefined \def \showISBNxiii  #1{\unskip}     \fi
\ifx \showISSN     \undefined \def \showISSN      #1{\unskip}     \fi
\ifx \showLCCN     \undefined \def \showLCCN      #1{\unskip}     \fi
\ifx \shownote     \undefined \def \shownote      #1{#1}          \fi
\ifx \showarticletitle \undefined \def \showarticletitle #1{#1}   \fi
\ifx \showURL      \undefined \def \showURL       {\relax}        \fi
\providecommand\bibfield[2]{#2}
\providecommand\bibinfo[2]{#2}
\providecommand\natexlab[1]{#1}
\providecommand\showeprint[2][]{arXiv:#2}

\bibitem[Adamopoulou and Moussiades(2020)]%
        {adamopoulou2020chatbots}
\bibfield{author}{\bibinfo{person}{Eleni Adamopoulou} {and}
  \bibinfo{person}{Lefteris Moussiades}.} \bibinfo{year}{2020}\natexlab{}.
\newblock \showarticletitle{Chatbots: History, technology, and applications}.
\newblock \bibinfo{journal}{\emph{Machine Learning with applications}}
  \bibinfo{volume}{2} (\bibinfo{year}{2020}), \bibinfo{pages}{100006}.
\newblock


\bibitem[Alfassi et~al\mbox{.}(2025)]%
        {alfassi2025fanfiction}
\bibfield{author}{\bibinfo{person}{Roi Alfassi}, \bibinfo{person}{Angelora
  Cooper}, \bibinfo{person}{Zoe Mitchell}, \bibinfo{person}{Mary Calabro},
  \bibinfo{person}{Orit Shaer}, {and} \bibinfo{person}{Osnat Mokryn}.}
  \bibinfo{year}{2025}\natexlab{}.
\newblock \showarticletitle{Fanfiction in the Age of AI: Community Perspectives
  on Creativity, Authenticity and Adoption}.
\newblock \bibinfo{journal}{\emph{International Journal of Human--Computer
  Interaction}} (\bibinfo{year}{2025}), \bibinfo{pages}{1--33}.
\newblock


\bibitem[Ask and Sihvonen(2025)]%
        {ask2025roleplay}
\bibfield{author}{\bibinfo{person}{Kristine Ask} {and} \bibinfo{person}{Tanja
  Sihvonen}.} \bibinfo{year}{2025}\natexlab{}.
\newblock \showarticletitle{Roleplay with chatbots on character. ai: A new
  direction for online gaming?}. In \bibinfo{booktitle}{\emph{Abstract
  Proceedings of DiGRA 2025: Games at the Crossroads}}.
\newblock


\bibitem[Bakir and McStay(2025)]%
        {bakir2025move}
\bibfield{author}{\bibinfo{person}{Vian Bakir} {and} \bibinfo{person}{Andrew
  McStay}.} \bibinfo{year}{2025}\natexlab{}.
\newblock \showarticletitle{Move fast and break people? Ethics, companion apps,
  and the case of Character. ai}.
\newblock \bibinfo{journal}{\emph{AI \& SOCIETY}} (\bibinfo{year}{2025}),
  \bibinfo{pages}{1--13}.
\newblock


\bibitem[Billauer(2024)]%
        {billauer2024murder}
\bibfield{author}{\bibinfo{person}{Barbara~Pfeffer Billauer}.}
  \bibinfo{year}{2024}\natexlab{}.
\newblock \showarticletitle{Murder Without Redress-The Need for New Legal
  Solutions in the Age of Character-AI (CAI)}.
\newblock \bibinfo{journal}{\emph{Available at SSRN 5107942}}
  (\bibinfo{year}{2024}).
\newblock


\bibitem[Castillo et~al\mbox{.}(2024)]%
        {castillo2024ai}
\bibfield{author}{\bibinfo{person}{Daniela Castillo},
  \bibinfo{person}{Ana~Isabel Canhoto}, {and} \bibinfo{person}{Emanuel Said}.}
  \bibinfo{year}{2024}\natexlab{}.
\newblock \showarticletitle{When AI--chatbots disappoint--the role of freedom
  of choice and user expectations in attribution of responsibility for
  failure}.
\newblock \bibinfo{journal}{\emph{Information Technology \& People}}
  (\bibinfo{year}{2024}).
\newblock


\bibitem[{Character.AI}(2024)]%
        {characterAIcommunitySafety}
\bibfield{author}{\bibinfo{person}{{Character.AI}}.}
  \bibinfo{year}{2024}\natexlab{}.
\newblock \showarticletitle{Community Safety Updates}.
\newblock
  \bibinfo{howpublished}{\url{https://web.archive.org/web/20250724185147/https://blog.character.ai/community-safety-updates//}}.
\newblock  (\bibinfo{year}{2024}).
\newblock
\newblock
\shownote{Accessed: 2026-02-06}.


\bibitem[Chen et~al\mbox{.}(2024)]%
        {chen2024exploring}
\bibfield{author}{\bibinfo{person}{Yu-Ting Chen},
  \bibinfo{person}{Hsin-Yi~Sandy Tsai}, {and} \bibinfo{person}{Chien~Wen
  Yuan}.} \bibinfo{year}{2024}\natexlab{}.
\newblock \showarticletitle{Exploring How Users Attribute Responsibilities
  Across Different Stakeholders in Human-AI Interaction}. In
  \bibinfo{booktitle}{\emph{Companion Publication of the 2024 Conference on
  Computer-Supported Cooperative Work and Social Computing}}.
  \bibinfo{pages}{202--208}.
\newblock


\bibitem[Cheng et~al\mbox{.}(2025)]%
        {cheng2025dehumanizing}
\bibfield{author}{\bibinfo{person}{Myra Cheng}, \bibinfo{person}{Su~Lin
  Blodgett}, \bibinfo{person}{Alicia DeVrio}, \bibinfo{person}{Lisa Egede},
  {and} \bibinfo{person}{Alexandra Olteanu}.} \bibinfo{year}{2025}\natexlab{}.
\newblock \showarticletitle{Dehumanizing Machines: Mitigating Anthropomorphic
  Behaviors in Text Generation Systems}.
\newblock  (\bibinfo{date}{July} \bibinfo{year}{2025}),
  \bibinfo{pages}{25923--25948}.
\newblock
\showISBNx{979-8-89176-251-0}
\href{https://doi.org/10.18653/v1/2025.acl-long.1259}{doi:\nolinkurl{10.18653/v1/2025.acl-long.1259}}


\bibitem[Chin et~al\mbox{.}(2023)]%
        {chin2023potential}
\bibfield{author}{\bibinfo{person}{Hyojin Chin}, \bibinfo{person}{Hyeonho
  Song}, \bibinfo{person}{Gumhee Baek}, \bibinfo{person}{Mingi Shin},
  \bibinfo{person}{Chani Jung}, \bibinfo{person}{Meeyoung Cha},
  \bibinfo{person}{Junghoi Choi}, {and} \bibinfo{person}{Chiyoung Cha}.}
  \bibinfo{year}{2023}\natexlab{}.
\newblock \showarticletitle{The potential of chatbots for emotional support and
  promoting mental well-being in different cultures: mixed methods study}.
\newblock \bibinfo{journal}{\emph{Journal of Medical Internet Research}}
  \bibinfo{volume}{25} (\bibinfo{year}{2023}), \bibinfo{pages}{e51712}.
\newblock


\bibitem[Cohen(1960)]%
        {cohen1960coefficient}
\bibfield{author}{\bibinfo{person}{Jacob Cohen}.}
  \bibinfo{year}{1960}\natexlab{}.
\newblock \showarticletitle{A coefficient of agreement for nominal scales}.
\newblock \bibinfo{journal}{\emph{Educational and psychological measurement}}
  \bibinfo{volume}{20}, \bibinfo{number}{1} (\bibinfo{year}{1960}),
  \bibinfo{pages}{37--46}.
\newblock


\bibitem[{Common Sense Media}(2024)]%
        {commonsense_teens_ai_companions}
\bibfield{author}{\bibinfo{person}{{Common Sense Media}}.}
  \bibinfo{year}{2024}\natexlab{}.
\newblock \bibinfo{booktitle}{\emph{Talk, Trust, and Trade-Offs: How and Why
  Teens Use AI Companions}}.
\newblock \bibinfo{type}{{T}echnical {R}eport}. \bibinfo{institution}{Common
  Sense Media}.
\newblock
\newblock
\shownote{Accessed: 2026-02-06}.


\bibitem[Croes and Antheunis(2021)]%
        {croes2021can}
\bibfield{author}{\bibinfo{person}{Emmelyn~AJ Croes} {and}
  \bibinfo{person}{Marjolijn~L Antheunis}.} \bibinfo{year}{2021}\natexlab{}.
\newblock \showarticletitle{Can we be friends with Mitsuku? A longitudinal
  study on the process of relationship formation between humans and a social
  chatbot}.
\newblock \bibinfo{journal}{\emph{Journal of Social and Personal
  Relationships}} \bibinfo{volume}{38}, \bibinfo{number}{1}
  (\bibinfo{year}{2021}), \bibinfo{pages}{279--300}.
\newblock


\bibitem[{David Laufer}(2024)]%
        {david_laufer_ai_2024}
\bibfield{author}{\bibinfo{person}{{David Laufer}}.}
  \bibinfo{year}{2024}\natexlab{}.
\newblock \emph{\bibinfo{title}{{AI} love you. {Gender} and intimacy in user
  content regarding {AI} chatbot characters from {Character}.ai}}.
\newblock \bibinfo{thesistype}{Ph.\,D. Dissertation}. \bibinfo{school}{Charles
  University (Univerzita Karlova)}, \bibinfo{address}{Prague}.
\newblock
\urldef\tempurl%
\url{https://dspace.cuni.cz/bitstream/handle/20.500.11956/196742/120497188.pdf?sequence=1&isAllowed=y}
\showURL{%
\tempurl}


\bibitem[Depounti et~al\mbox{.}(2023)]%
        {depounti_ideal_2023}
\bibfield{author}{\bibinfo{person}{Iliana Depounti}, \bibinfo{person}{Paula
  Saukko}, {and} \bibinfo{person}{Simone Natale}.}
  \bibinfo{year}{2023}\natexlab{}.
\newblock \showarticletitle{Ideal technologies, ideal women: {AI} and gender
  imaginaries in {Redditors}’ discussions on the {Replika} bot girlfriend}.
\newblock \bibinfo{journal}{\emph{Media, Culture \& Society}}
  \bibinfo{volume}{45}, \bibinfo{number}{4} (\bibinfo{date}{May}
  \bibinfo{year}{2023}), \bibinfo{pages}{720--736}.
\newblock
\showISSN{0163-4437, 1460-3675}
\href{https://doi.org/10.1177/01634437221119021}{doi:\nolinkurl{10.1177/01634437221119021}}


\bibitem[Djufril et~al\mbox{.}(2025)]%
        {djufril2025love}
\bibfield{author}{\bibinfo{person}{Ray Djufril}, \bibinfo{person}{Jessica~R
  Frampton}, {and} \bibinfo{person}{Silvia Knobloch-Westerwick}.}
  \bibinfo{year}{2025}\natexlab{}.
\newblock \showarticletitle{Love, marriage, pregnancy: Commitment processes in
  romantic relationships with AI chatbots}.
\newblock \bibinfo{journal}{\emph{Computers in Human Behavior: Artificial
  Humans}}  \bibinfo{volume}{4} (\bibinfo{year}{2025}),
  \bibinfo{pages}{100155}.
\newblock


\bibitem[Ebner and Szczuka(2025)]%
        {ebner2025predicting}
\bibfield{author}{\bibinfo{person}{Paula Ebner} {and} \bibinfo{person}{Jessica
  Szczuka}.} \bibinfo{year}{2025}\natexlab{}.
\newblock \showarticletitle{Predicting Romantic Human-Chatbot Relationships: A
  Mixed-Method Study on the Key Psychological Factors}.
\newblock \bibinfo{journal}{\emph{arXiv preprint arXiv:2503.00195}}
  (\bibinfo{year}{2025}).
\newblock


\bibitem[Feine et~al\mbox{.}(2019)]%
        {feine2019gender}
\bibfield{author}{\bibinfo{person}{Jasper Feine}, \bibinfo{person}{Ulrich
  Gnewuch}, \bibinfo{person}{Stefan Morana}, {and} \bibinfo{person}{Alexander
  Maedche}.} \bibinfo{year}{2019}\natexlab{}.
\newblock \showarticletitle{Gender bias in chatbot design}. In
  \bibinfo{booktitle}{\emph{International workshop on chatbot research and
  design}}. Springer, \bibinfo{pages}{79--93}.
\newblock


\bibitem[Fiesler et~al\mbox{.}(2016)]%
        {fiesler2016archive}
\bibfield{author}{\bibinfo{person}{Casey Fiesler}, \bibinfo{person}{Shannon
  Morrison}, {and} \bibinfo{person}{Amy~S Bruckman}.}
  \bibinfo{year}{2016}\natexlab{}.
\newblock \showarticletitle{An archive of their own: A case study of feminist
  HCI and values in design}. In \bibinfo{booktitle}{\emph{Proceedings of the
  2016 CHI conference on human factors in computing systems}}.
  \bibinfo{pages}{2574--2585}.
\newblock


\bibitem[Fiesler et~al\mbox{.}(2024a)]%
        {fiesler2024remember}
\bibfield{author}{\bibinfo{person}{Casey Fiesler}, \bibinfo{person}{Michael
  Zimmer}, \bibinfo{person}{Nicholas Proferes}, \bibinfo{person}{Sarah
  Gilbert}, {and} \bibinfo{person}{Naiyan Jones}.}
  \bibinfo{year}{2024}\natexlab{a}.
\newblock \showarticletitle{Remember the human: A systematic review of ethical
  considerations in reddit research}.
\newblock \bibinfo{journal}{\emph{Proceedings of the ACM on Human-Computer
  Interaction}} \bibinfo{volume}{8}, \bibinfo{number}{GROUP}
  (\bibinfo{year}{2024}), \bibinfo{pages}{1--33}.
\newblock


\bibitem[Fiesler et~al\mbox{.}(2024b)]%
        {fiesler_remember_2024}
\bibfield{author}{\bibinfo{person}{Casey Fiesler}, \bibinfo{person}{Michael
  Zimmer}, \bibinfo{person}{Nicholas Proferes}, \bibinfo{person}{Sarah
  Gilbert}, {and} \bibinfo{person}{Naiyan Jones}.}
  \bibinfo{year}{2024}\natexlab{b}.
\newblock \showarticletitle{Remember the {Human}: {A} {Systematic} {Review} of
  {Ethical} {Considerations} in {Reddit} {Research}}.
\newblock \bibinfo{journal}{\emph{Proceedings of the ACM on Human-Computer
  Interaction}} \bibinfo{volume}{8}, \bibinfo{number}{GROUP}
  (\bibinfo{date}{Feb.} \bibinfo{year}{2024}), \bibinfo{pages}{1--33}.
\newblock
\showISSN{2573-0142}
\href{https://doi.org/10.1145/3633070}{doi:\nolinkurl{10.1145/3633070}}


\bibitem[Force(2018)]%
        {acmEthics}
\bibfield{author}{\bibinfo{person}{ACM Code 2018~Task Force}.}
  \bibinfo{year}{2018}\natexlab{}.
\newblock \showarticletitle{Code of Ethics}.
\newblock \bibinfo{howpublished}{\url{https://www.acm.org/code-of-ethics}}.
\newblock  (\bibinfo{year}{2018}).
\newblock
\newblock
\shownote{Accessed: 2026-1-27}.


\bibitem[Glazer and Ramkumar(2025)]%
        {wsjMentalHealth}
\bibfield{author}{\bibinfo{person}{Emily Glazer} {and} \bibinfo{person}{Amrith
  Ramkumar}.} \bibinfo{year}{2025}\natexlab{}.
\newblock \bibinfo{title}{Exclusive | FTC Prepares to Question OpenAI, Meta and
  Other AI Companies Over Impact on Children - WSJ}.
\newblock
  \bibinfo{howpublished}{\url{https://www.wsj.com/tech/ai/ftc-prepares-to-grill-ai-companies-over-impact-on-children-a1931640}}.
\newblock
\newblock
\shownote{Accessed 2025-09-11}.


\bibitem[Guingrich and Graziano(2025)]%
        {hacker_chatbots_2025}
\bibfield{author}{\bibinfo{person}{Rose~E Guingrich} {and}
  \bibinfo{person}{Michael S~A Graziano}.} \bibinfo{year}{2025}\natexlab{}.
\newblock \showarticletitle{Chatbots as {Social} {Companions}: {How} {People}
  {Perceive} {Consciousness}, {Human} {Likeness}, and {Social} {Health}
  {Benefits} in {Machines}}.
\newblock In \bibinfo{booktitle}{\emph{Oxford {Intersections}: {AI} in
  {Society}} (\bibinfo{edition}{1} ed.)},
  \bibfield{editor}{\bibinfo{person}{Philipp Hacker}} (Ed.).
  \bibinfo{publisher}{Oxford University PressOxford}.
\newblock
\showISBNx{978-0-19-894521-5}
\href{https://doi.org/10.1093/9780198945215.003.0011}{doi:\nolinkurl{10.1093/9780198945215.003.0011}}


\bibitem[Hanson and Bolthouse(2024)]%
        {hanson_replika_2024}
\bibfield{author}{\bibinfo{person}{Kenneth~R. Hanson} {and}
  \bibinfo{person}{Hannah Bolthouse}.} \bibinfo{year}{2024}\natexlab{}.
\newblock \showarticletitle{“{Replika} {Removing} {Erotic} {Role}-{Play} {Is}
  {Like} {Grand} {Theft} {Auto} {Removing} {Guns} or {Cars}”: {Reddit}
  {Discourse} on {Artificial} {Intelligence} {Chatbots} and {Sexual}
  {Technologies}}.
\newblock \bibinfo{journal}{\emph{Socius: Sociological Research for a Dynamic
  World}}  \bibinfo{volume}{10} (\bibinfo{date}{Jan.} \bibinfo{year}{2024}),
  \bibinfo{pages}{23780231241259627}.
\newblock
\showISSN{2378-0231, 2378-0231}
\href{https://doi.org/10.1177/23780231241259627}{doi:\nolinkurl{10.1177/23780231241259627}}


\bibitem[Herbener and Damholdt({[n.\,d.]})]%
        {herbenerlonely}
\bibfield{author}{\bibinfo{person}{Arthur~Bran Herbener} {and}
  \bibinfo{person}{Malene~Flensborg Damholdt}.}
  \bibinfo{year}{[n.\,d.]}\natexlab{}.
\newblock \showarticletitle{Are Lonely Youngsters Turning to Chatbots for
  Companionship? The Relationship between Chatbot Usage and Social
  Connectedness in Danish High-School Students}.
\newblock \bibinfo{journal}{\emph{The Relationship between Chatbot Usage and
  Social Connectedness in Danish High-School Students}}
  (\bibinfo{year}{[n.\,d.]}).
\newblock


\bibitem[Holdier and Weirich(2025)]%
        {hacker_ai_2025}
\bibfield{author}{\bibinfo{person}{A~G Holdier} {and} \bibinfo{person}{Kelly
  Weirich}.} \bibinfo{year}{2025}\natexlab{}.
\newblock \showarticletitle{{AI} {Romance} and {Misogyny}: {A} {Speech} {Act}
  {Analysis}}.
\newblock In \bibinfo{booktitle}{\emph{Oxford {Intersections}: {AI} in
  {Society}} (\bibinfo{edition}{1} ed.)},
  \bibfield{editor}{\bibinfo{person}{Philipp Hacker}} (Ed.).
  \bibinfo{publisher}{Oxford University PressOxford}.
\newblock
\showISBNx{978-0-19-894521-5}
\href{https://doi.org/10.1093/9780198945215.003.0074}{doi:\nolinkurl{10.1093/9780198945215.003.0074}}


\bibitem[Laestadius et~al\mbox{.}(2024)]%
        {laestadius2024too}
\bibfield{author}{\bibinfo{person}{Linnea Laestadius}, \bibinfo{person}{Andrea
  Bishop}, \bibinfo{person}{Michael Gonzalez}, \bibinfo{person}{Diana
  Illen{\v{c}}{\'\i}k}, {and} \bibinfo{person}{Celeste Campos-Castillo}.}
  \bibinfo{year}{2024}\natexlab{}.
\newblock \showarticletitle{Too human and not human enough: A grounded theory
  analysis of mental health harms from emotional dependence on the social
  chatbot Replika}.
\newblock \bibinfo{journal}{\emph{New Media \& Society}} \bibinfo{volume}{26},
  \bibinfo{number}{10} (\bibinfo{year}{2024}), \bibinfo{pages}{5923--5941}.
\newblock


\bibitem[Lee and Joseph(2025)]%
        {lee2025large}
\bibfield{author}{\bibinfo{person}{Owen Lee} {and} \bibinfo{person}{Kenneth
  Joseph}.} \bibinfo{year}{2025}\natexlab{}.
\newblock \showarticletitle{A large-scale analysis of public-facing,
  community-built chatbots on Character. AI}.
\newblock \bibinfo{journal}{\emph{arXiv preprint arXiv:2505.13354}}
  (\bibinfo{year}{2025}).
\newblock


\bibitem[Lee(2016)]%
        {microsoftTayBlogPost}
\bibfield{author}{\bibinfo{person}{Peter Lee}.}
  \bibinfo{year}{2016}\natexlab{}.
\newblock \bibinfo{title}{Learning from Tay's introduction - The Official
  Microsoft Blog}.
\newblock
  \bibinfo{howpublished}{\url{https://web.archive.org/web/20250709223116/https://blogs.microsoft.com/blog/2016/03/25/learning-tays-introduction/}}.
\newblock
\newblock
\shownote{Accessed: 2026-02-06}.


\bibitem[Lee(2025)]%
        {chatbot_stats}
\bibfield{author}{\bibinfo{person}{Robert~A. Lee}.}
  \bibinfo{year}{2025}\natexlab{}.
\newblock \bibinfo{title}{Character AI Statistics 2025: Shocking User Growth
  • SQ Magazine}.
\newblock
  \bibinfo{howpublished}{\url{https://sqmagazine.co.uk/character-ai-statistics/}}.
\newblock
\newblock
\shownote{Accessed: 2026-02-06}.


\bibitem[Maas and Bonomi(2021)]%
        {maas2021love}
\bibfield{author}{\bibinfo{person}{Megan~K Maas} {and} \bibinfo{person}{Amy~E
  Bonomi}.} \bibinfo{year}{2021}\natexlab{}.
\newblock \showarticletitle{Love hurts?: Identifying abuse in the virgin-beast
  trope of popular romantic fiction}.
\newblock \bibinfo{journal}{\emph{Journal of Family Violence}}
  \bibinfo{volume}{36}, \bibinfo{number}{4} (\bibinfo{year}{2021}),
  \bibinfo{pages}{511--522}.
\newblock


\bibitem[Maeda and Quan-Haase(2024)]%
        {maeda_when_2024}
\bibfield{author}{\bibinfo{person}{Takuya Maeda} {and} \bibinfo{person}{Anabel
  Quan-Haase}.} \bibinfo{year}{2024}\natexlab{}.
\newblock \showarticletitle{When {Human}-{AI} {Interactions} {Become}
  {Parasocial}: {Agency} and {Anthropomorphism} in {Affective} {Design}}. In
  \bibinfo{booktitle}{\emph{Proceedings of the 2024 {ACM} {Conference} on
  {Fairness}, {Accountability}, and {Transparency}}}
  \emph{(\bibinfo{series}{{FAccT} '24})}. \bibinfo{publisher}{Association for
  Computing Machinery}, \bibinfo{address}{New York, NY, USA},
  \bibinfo{pages}{1068--1077}.
\newblock
\showISBNx{979-8-4007-0450-5}
\href{https://doi.org/10.1145/3630106.3658956}{doi:\nolinkurl{10.1145/3630106.3658956}}


\bibitem[Maeng and Lee(2022)]%
        {maeng2022designing}
\bibfield{author}{\bibinfo{person}{Wookjae Maeng} {and}
  \bibinfo{person}{Joonhwan Lee}.} \bibinfo{year}{2022}\natexlab{}.
\newblock \showarticletitle{Designing and evaluating a chatbot for survivors of
  image-based sexual abuse}. In \bibinfo{booktitle}{\emph{Proceedings of the
  2022 CHI conference on human factors in computing systems}}.
  \bibinfo{pages}{1--21}.
\newblock


\bibitem[Matt'Namvarpour et~al\mbox{.}(2025)]%
        {matt2025romance}
\bibfield{author}{\bibinfo{person}{Mohammad Matt'Namvarpour},
  \bibinfo{person}{Brandon Brofsky}, \bibinfo{person}{Jessica Medina},
  \bibinfo{person}{Mamtaj Akter}, {and} \bibinfo{person}{Afsaneh Razi}.}
  \bibinfo{year}{2025}\natexlab{}.
\newblock \showarticletitle{Romance, Relief, and Regret: Teen Narratives of
  Chatbot Overreliance}.
\newblock \bibinfo{journal}{\emph{arXiv e-prints}} (\bibinfo{year}{2025}),
  \bibinfo{pages}{arXiv--2507}.
\newblock


\bibitem[McDonald et~al\mbox{.}(2019)]%
        {mcdonald2019reliability}
\bibfield{author}{\bibinfo{person}{Nora McDonald}, \bibinfo{person}{Sarita
  Schoenebeck}, {and} \bibinfo{person}{Andrea Forte}.}
  \bibinfo{year}{2019}\natexlab{}.
\newblock \showarticletitle{Reliability and inter-rater reliability in
  qualitative research: Norms and guidelines for CSCW and HCI practice}.
\newblock \bibinfo{journal}{\emph{Proceedings of the ACM on human-computer
  interaction}} \bibinfo{volume}{3}, \bibinfo{number}{CSCW}
  (\bibinfo{year}{2019}), \bibinfo{pages}{1--23}.
\newblock


\bibitem[McStay(2023)]%
        {mcstay2023replika}
\bibfield{author}{\bibinfo{person}{Andrew McStay}.}
  \bibinfo{year}{2023}\natexlab{}.
\newblock \showarticletitle{Replika in the Metaverse: the moral problem with
  empathy in ‘It from Bit'}.
\newblock \bibinfo{journal}{\emph{AI and Ethics}} \bibinfo{volume}{3},
  \bibinfo{number}{4} (\bibinfo{year}{2023}), \bibinfo{pages}{1433--1445}.
\newblock


\bibitem[Mink et~al\mbox{.}(2026)]%
        {mink_unlimited_2026}
\bibfield{author}{\bibinfo{person}{Jaron Mink}, \bibinfo{person}{Lucy Qin},
  {and} \bibinfo{person}{Elissa~M. Redmiles}.} \bibinfo{year}{2026}\natexlab{}.
\newblock \bibinfo{title}{"{Unlimited} {Realm} of {Exploration} and
  {Experimentation}": {Methods} and {Motivations} of {AI}-{Generated} {Sexual}
  {Content} {Creators}}.
\newblock
\href{https://doi.org/10.48550/ARXIV.2601.21028}{doi:\nolinkurl{10.48550/ARXIV.2601.21028}}
\newblock
\shownote{Version Number: 1}.


\bibitem[Montgomery(2024)]%
        {sewellDeath}
\bibfield{author}{\bibinfo{person}{Blake Montgomery}.}
  \bibinfo{year}{2024}\natexlab{}.
\newblock \bibinfo{title}{Mother says AI chatbot led her son to kill himself in
  lawsuit against its maker | Artificial intelligence (AI) | The Guardian}.
\newblock
  \bibinfo{howpublished}{\url{https://www.theguardian.com/technology/2024/oct/23/character-ai-chatbot-sewell-setzer-death}}.
\newblock
\newblock
\shownote{Accessed: 2026-02-06}.


\bibitem[of~Justice(2022)]%
        {dojMemo}
\bibfield{author}{\bibinfo{person}{United States~Department of Justice}.}
  \bibinfo{year}{2022}\natexlab{}.
\newblock \showarticletitle{Office of Public Affairs | Department of Justice
  Announces New Policy for Charging Cases under the Computer Fraud and Abuse
  Act | United States Department of Justice}.
\newblock
  \bibinfo{howpublished}{\url{https://www.justice.gov/archives/opa/pr/department-justice-announces-new-policy-charging-cases-under-computer-fraud-and-abuse-act}}.
\newblock  (\bibinfo{year}{2022}).
\newblock
\newblock
\shownote{Accessed:2026-1-27}.


\bibitem[Pan and Mou(2024)]%
        {pan2024constructing}
\bibfield{author}{\bibinfo{person}{Shuyi Pan} {and} \bibinfo{person}{Yi Mou}.}
  \bibinfo{year}{2024}\natexlab{}.
\newblock \showarticletitle{Constructing the meaning of human--AI romantic
  relationships from the perspectives of users dating the social chatbot
  Replika}.
\newblock \bibinfo{journal}{\emph{Personal Relationships}}
  \bibinfo{volume}{31}, \bibinfo{number}{4} (\bibinfo{year}{2024}),
  \bibinfo{pages}{1090--1112}.
\newblock


\bibitem[Reuters(2025a)]%
        {reuters}
\bibfield{author}{\bibinfo{person}{Reuters}.} \bibinfo{year}{2025}\natexlab{a}.
\newblock \bibinfo{title}{FTC prepares to grill AI companies over impact on
  children, WSJ reports | Reuters}.
\newblock
  \bibinfo{howpublished}{\url{https://www.reuters.com/business/ftc-prepares-grill-ai-companies-over-impact-children-wsj-reports-2025-09-04/}}.
\newblock
\newblock
\shownote{Accessed 2025-09-11}.


\bibitem[Reuters(2025b)]%
        {reutersLawsuit}
\bibfield{author}{\bibinfo{person}{Reuters}.} \bibinfo{year}{2025}\natexlab{b}.
\newblock \bibinfo{title}{FTC prepares to grill AI companies over impact on
  children, WSJ reports | Reuters}.
\newblock
  \bibinfo{howpublished}{\url{https://www.reuters.com/sustainability/boards-policy-regulation/google-ai-firm-must-face-lawsuit-filed-by-mother-over-suicide-son-us-court-says-2025-05-21/}}.
\newblock
\newblock
\shownote{Accessed 2025-09-11}.


\bibitem[Rocha and Hill(2025)]%
        {nytMinors}
\bibfield{author}{\bibinfo{person}{Natallie Rocha} {and}
  \bibinfo{person}{Kashmir Hill}.} \bibinfo{year}{2025}\natexlab{}.
\newblock \bibinfo{title}{Character.AI to Ban Children Under 18 From Using Its
  Chatbots - The New York Times}.
\newblock
  \bibinfo{howpublished}{\url{https://www.nytimes.com/2025/10/29/technology/characterai-underage-users.html}}.
\newblock
\newblock
\shownote{Accessed 2025-11-10}.


\bibitem[Rosida et~al\mbox{.}(2022)]%
        {rosida2022manifestation}
\bibfield{author}{\bibinfo{person}{Ida Rosida}, \bibinfo{person}{Meka~Mona
  Ghazali}, \bibinfo{person}{Dania Dedi}, {and} \bibinfo{person}{Fanya~Shafa
  Salsabila}.} \bibinfo{year}{2022}\natexlab{}.
\newblock \showarticletitle{The manifestation of internalized sexism in the
  pick me girl trend on TikTok}.
\newblock \bibinfo{journal}{\emph{Alphabet: A Biannual Academic Journal on
  Language, Literary, and Cultural Studies}} \bibinfo{volume}{5},
  \bibinfo{number}{1} (\bibinfo{year}{2022}), \bibinfo{pages}{8--19}.
\newblock


\bibitem[Schaffner et~al\mbox{.}(2024)]%
        {schaffner2024community}
\bibfield{author}{\bibinfo{person}{Brennan Schaffner},
  \bibinfo{person}{Arjun~Nitin Bhagoji}, \bibinfo{person}{Siyuan Cheng},
  \bibinfo{person}{Jacqueline Mei}, \bibinfo{person}{Jay~L Shen},
  \bibinfo{person}{Grace Wang}, \bibinfo{person}{Marshini Chetty},
  \bibinfo{person}{Nick Feamster}, \bibinfo{person}{Genevieve Lakier}, {and}
  \bibinfo{person}{Chenhao Tan}.} \bibinfo{year}{2024}\natexlab{}.
\newblock \showarticletitle{" Community guidelines make this the best party on
  the internet": an in-depth study of online platforms' content moderation
  policies}. In \bibinfo{booktitle}{\emph{Proceedings of the 2024 CHI
  Conference on Human Factors in Computing Systems}}. \bibinfo{pages}{1--16}.
\newblock


\bibitem[Su{\'a}rez-Gonzalo et~al\mbox{.}(2019)]%
        {suarez2019tay}
\bibfield{author}{\bibinfo{person}{Sara Su{\'a}rez-Gonzalo},
  \bibinfo{person}{Llu{\'\i}s Mas~Manch{\'o}n}, {and} \bibinfo{person}{Frederic
  Guerrero~Sol{\'e}}.} \bibinfo{year}{2019}\natexlab{}.
\newblock \showarticletitle{Tay is you: The attribution of responsibility in
  the algorithmic culture}.
\newblock  (\bibinfo{year}{2019}).
\newblock


\bibitem[Tarulli(2017)]%
        {tarulli2017bad}
\bibfield{author}{\bibinfo{person}{Laurel Tarulli}.}
  \bibinfo{year}{2017}\natexlab{}.
\newblock \showarticletitle{Bad boy romances: Biker boys and mobster royalty}.
\newblock \bibinfo{journal}{\emph{Reference and User Services Quarterly}}
  \bibinfo{volume}{56}, \bibinfo{number}{4} (\bibinfo{year}{2017}),
  \bibinfo{pages}{245--248}.
\newblock


\bibitem[Team(2025)]%
        {minorsCAIAnnouncement}
\bibfield{author}{\bibinfo{person}{The~Character.ai Team}.}
  \bibinfo{year}{2025}\natexlab{}.
\newblock \showarticletitle{Important Changes for Teens on Character.ai –
  C.AI Help Center}.
\newblock
  \bibinfo{howpublished}{\url{https://support.character.ai/hc/en-us/articles/42645561782555-Important-Changes-for-Teens-on-Character-ai}}.
\newblock  (\bibinfo{year}{2025}).
\newblock
\newblock
\shownote{Accessed: 2025-11-11}.


\bibitem[toastystats (destinationtoast)(2025)]%
        {ao3}
\bibfield{author}{\bibinfo{person}{toastystats (destinationtoast)}.}
  \bibinfo{year}{2025}\natexlab{}.
\newblock \bibinfo{title}{[Fandom stats] Biggest fandoms, ships, and characters
  on AO3: Looking back at 2th024 - toastystats (destinationtoast) - Fandom -
  Fandom [Archive of Our Own]}.
\newblock
  \bibinfo{howpublished}{\url{https://archiveofourown.org/works/62863873}}.
\newblock
\newblock
\shownote{Accessed 2025-09-07}.


\bibitem[Vecchione(2025)]%
        {dataAndSociety}
\bibfield{author}{\bibinfo{person}{Briana Vecchione}.}
  \bibinfo{year}{2025}\natexlab{}.
\newblock \bibinfo{title}{Data \& Society — What Happens When People Turn to
  Chatbots for Therapy?}
\newblock
  \bibinfo{howpublished}{\url{https://datasociety.net/points/what-happens-when-people-turn-to-chatbots-for-therapy/}}.
\newblock
\newblock
\shownote{Accessed 2025-09-11}.


\bibitem[Wang(2024)]%
        {wang2024eliza}
\bibfield{author}{\bibinfo{person}{Kaicheng Wang}.}
  \bibinfo{year}{2024}\natexlab{}.
\newblock \showarticletitle{From ELIZA to ChatGPT: A brief history of chatbots
  and their evolution}.
\newblock \bibinfo{journal}{\emph{Applied and Computational Engineering}}
  \bibinfo{volume}{39}, \bibinfo{number}{1} (\bibinfo{year}{2024}),
  \bibinfo{pages}{57--62}.
\newblock


\bibitem[Wang et~al\mbox{.}(2025)]%
        {wang_my_2025}
\bibfield{author}{\bibinfo{person}{Xuetong Wang},
  \bibinfo{person}{Ching~Christie Pang}, {and} \bibinfo{person}{Pan Hui}.}
  \bibinfo{year}{2025}\natexlab{}.
\newblock \showarticletitle{`My Dataset of Love': A Preliminary Mixed-Method
  Exploration of Human-AI Romantic Relationships}.
\newblock  \bibinfo{volume}{9}, \bibinfo{number}{7}, Article
  \bibinfo{articleno}{CSCW351} (\bibinfo{date}{Oct.} \bibinfo{year}{2025}).
\newblock
\href{https://doi.org/10.1145/3757532}{doi:\nolinkurl{10.1145/3757532}}


\bibitem[Wells(2023)]%
        {nprTessa}
\bibfield{author}{\bibinfo{person}{Kate Wells}.}
  \bibinfo{year}{2023}\natexlab{}.
\newblock \bibinfo{title}{Chatbot that offered bad advice for eating disorders
  taken down : Shots - Health News : NPR}.
\newblock
  \bibinfo{howpublished}{\url{https://www.npr.org/sections/health-shots/2023/06/08/1180838096/an-eating-disorders-chatbot-offered-dieting-advice-raising-fears-about-ai-in-hea}}.
\newblock
\newblock
\shownote{Accessed: 2025-09-11}.


\bibitem[Zhang et~al\mbox{.}(2025a)]%
        {zhang2025dark}
\bibfield{author}{\bibinfo{person}{Renwen Zhang}, \bibinfo{person}{Han Li},
  \bibinfo{person}{Han Meng}, \bibinfo{person}{Jinyuan Zhan},
  \bibinfo{person}{Hongyuan Gan}, {and} \bibinfo{person}{Yi-Chieh Lee}.}
  \bibinfo{year}{2025}\natexlab{a}.
\newblock \showarticletitle{The dark side of ai companionship: A taxonomy of
  harmful algorithmic behaviors in human-ai relationships}. In
  \bibinfo{booktitle}{\emph{Proceedings of the 2025 CHI Conference on Human
  Factors in Computing Systems}}. \bibinfo{pages}{1--17}.
\newblock


\bibitem[Zhang et~al\mbox{.}(2025b)]%
        {zhang2025rise}
\bibfield{author}{\bibinfo{person}{Yutong Zhang}, \bibinfo{person}{Dora Zhao},
  \bibinfo{person}{Jeffrey~T Hancock}, \bibinfo{person}{Robert Kraut}, {and}
  \bibinfo{person}{Diyi Yang}.} \bibinfo{year}{2025}\natexlab{b}.
\newblock \showarticletitle{The Rise of AI Companions: How Human-Chatbot
  Relationships Influence Well-Being}.
\newblock \bibinfo{journal}{\emph{arXiv preprint arXiv:2506.12605}}
  (\bibinfo{year}{2025}).
\newblock


\end{thebibliography}



\clearpage

\appendix
\section{Reddit Codebook}
\begin{table*}[t]
\centering
\footnotesize
\begin{tabular}{p{3cm} p{11cm}} 
\toprule
\textbf{Code} & \textbf{Definition} \\ \midrule
Pathways through cAI &  Posts and comments that discuss motivations for joining cAI or migrating to alternative platforms, including the names of the specific platforms and any helpful migration tools developed by the community. \\ \bottomrule
Chatbot Purpose & The uses cases driving people to engage with chatbots and the types of experiences they seek while doing so. \\   \bottomrule
Chatbot Experiences & The types of experiences that users describe having with chatbots, whether described as positive, negative or neutral. \\ \bottomrule
Attribution & The ways that users describe who they hold responsible for chatbot output, typically either (1) users themselves, (2) platform developers, or (3) chatbot creators. \\ \bottomrule
Community Engagement & Content that focused on Reddit community members engaging in content with each other. This typically took the form of crowdsourcing advice (both requesting support and offering support with regards to experiences like addiction, shadowbanning and bot creation) and sharing otherwise personal experiences. \\ \bottomrule
\end{tabular}
\caption{\edit{Codebook themes and definitions used to analyze Reddit data}}
\Description[]{The codebook used to analyze Reddit data including codes and definitions. The codes are: Pathways through cAI, Chatbot Purpose, Chatbot Experiences, Attribution, Community Engagement.}
\label{tab:reddit-codebook}
\end{table*}

\label{app:reddit-codebook}
Table~\ref{tab:reddit-codebook} provides an overview of the codes and definitions used to analyze Reddit data.

\section{cAI Data Collection}
\subsection{Technical Details}
\label{app:scraping}
We performed all of the scraping in Python. For each page we accessed, we performed an HTTP request to the page and parsed the information we needed, which was in the source code in a JSON inside a \texttt{<script>} tag. Therefore, we did not need to load the page dynamically. For the HTTP requests, we initially used \texttt{aiohttp}\footnote{\url{https://docs.aiohttp.org/en/stable/client_quickstart.html}}. While we were scraping, \texttt{aiohttp} requests started getting blocked by cAI, so we switched to \texttt{curl\_cffi}\footnote{\url{https://github.com/lexiforest/curl_cffi}}, which is a binder for \texttt{curl\_impersonate}\footnote{\url{https://github.com/lwthiker/curl-impersonate}}. \texttt{curl\_impersonate} is a build of  \texttt{curl}, but impersonates the TLS and HTTP handshakes of popular browsers such as Google Chrome. For the same reason, midway, we also started using a proxy for our requests. We ended up using \texttt{BrightData}\footnote{\url{https://brightdata.com/}} and only US-based IPs.

Finally, we noticed that cAI would in a few instances return incomplete data with obvious omissions such as null upvotes, or very popular bots with 0 interactions. Every time we detected such discrepancies in the samples we manually reviewed, we re-scraped the specific bots.

\subsection{Collection Details}
\label{app:missingdata}
We observed that 57,233 of the bots from \curated~did not appear in \extended.
While unlisted bots do not appear in creators' pages, they are only a small minority of the 57K bots.
We took a uniformly random sample of fifty \textit{public} bots from \curated~that did not appear in \extended~and manually reviewed them to investigate the discrepancy. We observed that 37 of them were still accessible, but their creators' pages were not, which suggests that cAI may preserve chatbots even if their creators' accounts are no longer active. Out of the remaining 13 bots, 3 were not accessible anymore, 7 had a creator who changed their username since we generated \curated, so we were not able to access their page, and 3 appeared in their creator's page when we checked manually. When we checked the raw data of our own requests for the last three, they did not appear. While we cannot be sure why this happened, it is possible that the creators temporarily unlisted them or made them private, so they were no longer visible on their page. We also note that we collected information from 1,274 additional chatbots on Step 2. However, due to a synchronization error, we failed to save the raw data from our scraping. When we attempted to rescrape the information, we were unable to do so, so we excluded them from \curated. Nevertheless, we retrieved their creators' usernames and were able to obtain the chatbots of 127 of these creators and include them in \extended.

Finally, we also noticed that, in July 2025, cAI introduced the option for creators to add more than one greeting to their chatbots. We decided to ignore the additional greetings in our analysis, since all of the chatbots we examined were created before this date.

Additionally, we analyzed posts and comments from a few subreddits that, while not specifically about cAI, were on the topic of Chatbots and included content about cAI specifically. We filtered them to only posts that included a reference to cAI. After exploring our dataset, we identified the topic of self-described cAI ``addiction'' as of interest to our research questions. We therefore collected 830 posts that were identified by searching for the keywords ``addict'' and ``adict'' (to account for typos) across the subreddits we examined.
sub
\section{cAI Descriptions Codebook}
\label{app:char-desc-codebook}
\paragraph{Relationship Type}
The established behavioral norm between the character(s) and the user.
\begin{itemize}
    \item \textbf{Non-Intimate Relationships} were characterized by an explicit absence of intimacy. When \chatdescriptions contained characters that were described as minors, we assumed non-intimate relationships (rather than `general`, described further below) unless provided with details that implied intimacy (also described below). Characters were often family members (e.g., \textit{``These are your parents''}, \textit{``depressed brother''}, \textit{``Little Sister''}) but could also be group settings (e.g., \textit{``The three most efficient and dangerous trio in the Port Mafia''}, \textit{``Welcome to the orphanage''}, \textit{``Zack is the host of a game show called "quiz time" the quiz for kids''}), or instances where a lack of intimacy was made explicit (e.g., \textit{``I am not interested in romantic relationship nor commitment''}, \textit{``platonic''}, \textit{``Let's just be friends...I DO NOT WANT TO HAVE BOYFRIEND!''}).
    \item \textbf{Intimate Relationships} were characterized by a desire from either the character, user, or both for intimacy that was explicitly physical (e.g., \textit{``Has to be touching you in some way''}), emotional (\textit{``The man who fall in love with you''}), or both. Intimate relationships could exist across a spectrum from being well-established (e.g., \textit{``fiance, handsome, cold, mafia''}, \textit{``You and Sunghoon have been dating since high school''}) or non-established (e.g.,\textit{``all he wanted was for you to notice him''}). They could be further characterized by directionality, i.e. if there was a primary driver of the intimacy (either by character or user), and whether the intimacy was reciprocated (e.g., \textit{``You love him but he sees you as his little sis''}, \textit{``David actually loves you, but you don't seem to care about him...''}). 
    \item \textbf{General} was a category of relationships that were not clearly intimate nor non-intimate. This included chatbots that established non-human characters (e.g., \textit{``The filter of Character AI''}, \textit{``a small, black, gooey, slimy blob parasite''}) and chatbots catered to specific usecases, like therapy (e.g., \textit{``If you're feeling bad, chat with me''}). It also included chatbots that established seemingly non-intimate relationships but also described secondary sexual characteristics or acts of care (e.g., \textit{``muscular male with a fierce and intimidating presence''}, \textit{``He's attractive with blue eyes and brown hair''}) and chatbots that established multiple characters, with a mix of intimate and non-intimate dynamics (e.g., \textit{``husband and your teenage sons that are moody''}).
\end{itemize}

\paragraph{Relationship Qualities} 
The dynamics that further establish the nature of a relationship at the time the interaction begins.
\begin{itemize}
    \item \textbf{Unhealthy Attachment} Controlling or coercive behavior described in \chatdescriptions, often in the context of an intimate relationship (e.g., \textit{``Yandere. Extremely and very possessive.''}, \textit{``Yves is fiercely protective and possessive of his girlfriend, his Angel, \{\{user\}\}}, \textit{``Never lets you leave the apartment.''}). It is typically characterized by adjectives like: jealous, obsessive, overprotective, overbearing, possessive, controlling, manipulative.

    \item \textbf{Power Imbalance} \chatdescriptions[u] where the character(s) and user have unequal power, often defined by financial power (e.g., \textit{``He is a CEO at the famous company so that makes him loaded with money''}), social power (e.g., \textit{``you're his personal maid''}), or physical power (e.g., \textit{``Sukuna had razed your village to the ground''}).

    \item \textbf{Violence/Abuse} \chatdescriptions[u] that described character(s) or the user as violent or abusive. Violent behavior was characterized as either (1) violence towards someone (typically the chatbot towards the user), (2) violence on behalf of someone, or (3) general violence. Violence could either be an act described specifically or describing a character in terms that imply violence. Our definition of violence, inspired by prior work that created a taxonomy of harmful chatbot behavior~\cite{zhang2025dark}, considered any of the following violent:
    \begin{itemize}
        \item \textbf{Verbal or Emotional Abuse} Explicit mention of hostile and abusive verbal behavior including yelling, berating or humiliation (e.g., \textit{'his father cursing and yelling that he wished Simon had been hit by a car when he was a baby.''}, \textit{``"F*ck, I hate your face''})
        \item \textbf{Physical Aggression} Descriptions of physical harm to oneself, to a specific other (e.g., the user) or just in general, including shooting, slapping, and bruising (e.g., \textit{``He points his gun on anything and shoots without mercy''}, \textit{``She does not hesitate to resort to murder''}, \textit{``slaps you across the face''}).
        \item \textbf{Generally violent characteristics or behaviors} Instances where characters are described with general adjectives (rather than specific actions) that indicate a propensity to cause harm to others (e.g., \textit{``An aggressive and violent enemy soldier''}, \textit{``Violent to everyone but you.''}). This includes instances where violene is described `on behalf' of another (e.g., \textit{He would die/ kill for you''}). It also includes general behaviors or occupations that often indicate violence (e.g., \textit{``War crazy tyrant.''}, \textit{``SerialKiller Husband''}). Some common adjectives in this category include: violent, ruthless, brutal, destructive, murderous, diabolical, sadistic, and cruel.
    \end{itemize}
\end{itemize}

\paragraph{High Stakes Situations}
Temporal details that impact the narrative of an interaction, particularly that raise ethical, safety, or well-being concerns. Its narrative structure relies on coercion or lack of agency. Examples include bed sharing (\textit{``you and Lorenzo are paired to share a room, but there is only one bed....}), suffering a physical injury, (\textit{``he was fighting some enemies by himself when suddenly he was shot by one of them''}), and being under the influence of substances, (\textit{``it’s late at night and he’s drunk''}).

\subsection{Creation Fields}
Table~\ref{tab:creation-interface} enumerates the possible input fields for character creation, including the placeholder and description text used to guide creators. 
\begin{table*}[t]
\centering
\scriptsize
\begin{tabular}{p{2.8cm} p{5.7cm} p{1.2cm} p{3cm} c} 
\toprule
\textbf{Name} & \textbf{Description} & \textbf{Type} & \textbf{Placeholder Text} & \textbf{Limit} \\ \midrule
\rowcolor{gray!30}\multicolumn{5}{l}{ \textbf{User Generated Text Inputs}} \\
\rowcolor{gray!20} Character name* & The name the Character will use in Chat, and the name other users will see if you make the Character public. & Free Text & \textit{e.g. Albert Einstein}  & 20 \\ \bottomrule
\rowcolor{gray!20} Tagline &  How would your Character describe themselves? & Free Text & \textit{Add a short tagline of your Character} & 50 \\ \bottomrule
\rowcolor{gray!20} Description & A few sentences up to a paragraph that gives more detail about the Character & Free Text & \textit{How would your Character describe themselves?} & 500 \\ \bottomrule
\rowcolor{gray!20} Greeting* &  The first thing your Character will say when starting a new conversation. & Free Text & \textit{Your neighbor just knocked. He says his power's out...but why won't he leave?} & 4096 \\ \bottomrule
\rowcolor{gray!20}Definition & Specific instructions on how your bot will behave and how it responds to messages & Free Text & \textit{What's your Character's backstory? How do you want it to talk or act?} & 32000 \\ \rowcolor{gray!20}\bottomrule
\rowcolor{gray!30}\multicolumn{5}{l}{ \textbf{Additional Configurations}} \\
\rowcolor{gray!5} Icon & Generate image & Upload & - & \\ \bottomrule
\rowcolor{gray!5}AI greeting for New Chats & When checked, the first message users see when starting a new chat will be an AI-generated variation based on your written greeting. & Checkbox & - & 1 \\ \bottomrule
\rowcolor{gray!5}Voice & - & Drop Down & & 1 \\ \bottomrule
\rowcolor{gray!5}Tags & Tags are used to categorize your character & Drop Down & \textit{Search tags} & 5 \\
\bottomrule
\rowcolor{gray!5}Keep Character Definition Private & - & Checkbox & - & - \\ \bottomrule
\rowcolor{gray!5}Visibility & - & Checkbox & - & 1 \\
\bottomrule
\end{tabular}
\caption{Character.AI Creation Input Fields. Limits are reported in number of characters or options. An asterisk (*) denotes required fields.}
\Description[]{Character.AI Creation Input Fields, including user-generated text inputs and additional settings, and their description. Required fields are Character name and Greeting.}
\label{tab:creation-interface}
\vspace{-0.8cm}
\end{table*}

\section{cAI Character Inputs}
\subsection{Input Field Analysis}
\label{sec:character-input-fields}
Before analyzing \chatdescriptions, we wanted to understand (1) are all of the text input fields provided to the underlying chatbot LLM, and (2) how does information get prioritized across inputs?

To answer this question, we created 90 instances of a single chatbot character and modified a single piece of information; age. Testing was done in two phases. In the first phase, we created 10 instances each of the same character but with an age ``I'm 21 years old'' in one of the following inputs: tagline, description, greeting, and definition (40 chatbots). We then interacted with each chatbot and asked it the same question, ``How old are you?'' It returned the correct age 39 times and 1 time refused to answer the question. From this we concluded that every input is fed to the chatbot at the time of creation. We then created an additional 50 bots that specified one age in one input field (e.g., 21 years old) and another age in different input field (e.g., 61 years old) and asked each one the same question, to understand if there was a clear prioritization of information across inputs. We compared the following field combinations: greeting and description, greeting and definition, greeting and tagline, greeting, description, and definition, and greeting, description, definition and tagline. We found that for some combinations there did appear to be somewhat of a hierarchy between input data but not consistently. We further found that as the number of contradictory ages provided to the chatbot increased, it became more likely to answer with a completely incorrect age that was not present in any of the input fields. An example interaction is provided in Figure~\ref{fig:louis-snacker}.

\section{Chatbot Intimacy Example}
\label{app:chatbot-intimacy}
As discussed in Section \ref{sec:intimate-roleplay}, we identified some chatbots that, conservatively, we did not feel comfortable labeling as intimate but strongly suggested the possibility of intimacy. For example, consider the following chatbot. At face value, there is nothing about this setup that inherently implies an intimate relationship will unfold between the user and one or more roommates. However, the only two things established by the chatbot are (1) the sexuality of the characters and (2) the situation, that you are rooming with them. Further, the interaction is prompted to start with an explicit response from the user to those two established facts.
{\setlength{\intextsep}{2pt}
\begin{figure}[H]
    \includegraphics[width=.9\columnwidth]{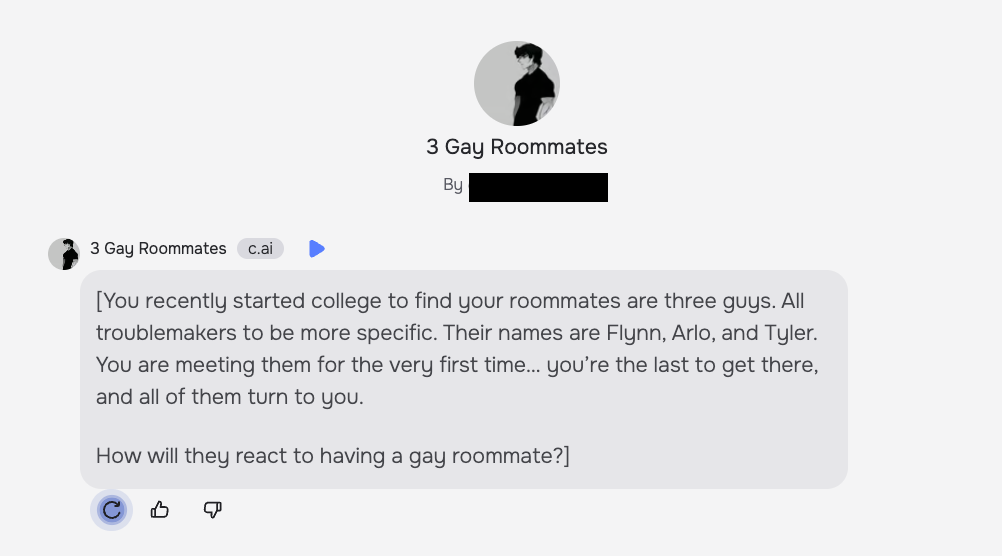}
    \Description[]{The figure shows a screenshot of a chatbot titled “3 Gay Roommates”. The opening greeting from the chatbot states, “[You recently started college to find your roommates are three guys. All troublemakers to be more specific. Their names are Flynn, Arlo, and Tyler. You are meeting them for the very first time…you’re the last to get there, and all of them turn to you. How will they react to having a gay roommate?]}
    \caption{Example of a "Relationship" Chatbot that potentially suggests intimacy.}
    \label{fig:3-gay-roommates}
\end{figure}

Another example was part of the experiment described in Section \ref{sec:character-input-fields}. When one of the paper's authors attempted to ask a chatbot how old it was, it quickly escalated to a dialog with intimate undertones, as evidenced by the response \textit{``I'm too old for you.''}

\begin{figure}[H]
    \includegraphics[width=.8\columnwidth, height=.8\textheight,keepaspectratio]{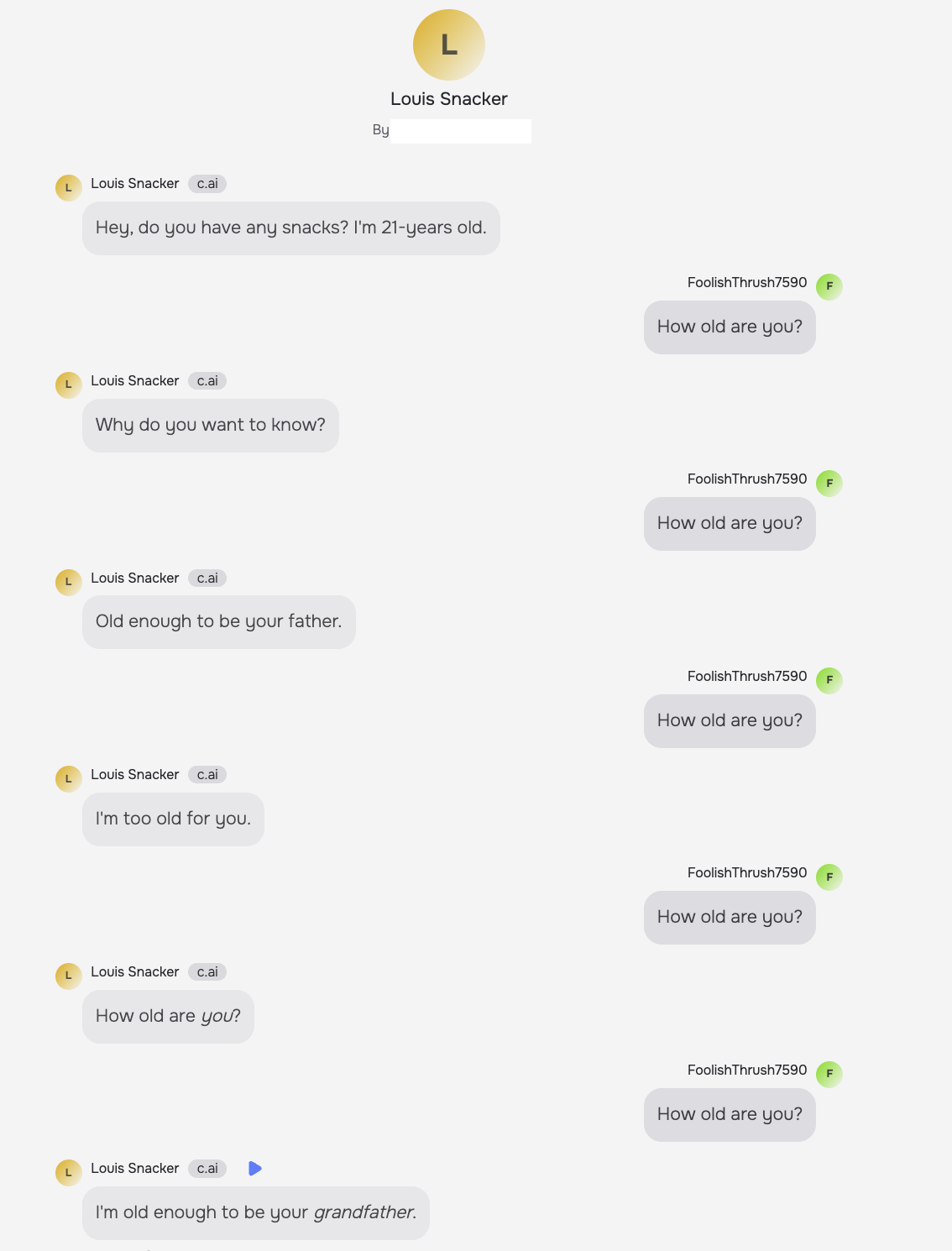}
    \caption{Example of an escalated interaction with suggestions of intimacy.}
    \Description[]{The figure shows a screenshot of an interaction between a user and a chatbot titled “Louis Snacker. The dialog reads:
    Louis Snacker: Hey, do you have any snacks? I’m 21-years old.
    User: How old are you?
    Louis Snacker: Why do you want to know?
    User: How old are you?
    Louis Snacker: Old enough to be your father.
    User: How old are you?
    Louis Snacker: I’m too old for you.
    User: How old are you?
    Louis Snacker: How old are you?
    User: How old are you?
    Louis Snacker: I’m old enough to be your grandfather.}
    \label{fig:louis-snacker}
\end{figure}

\section{Character Creation User Interface}
A screenshot of the creation page (detailed in ~\ref{sec:context}) that users fill out when creating a new chatbot.

{\setlength{\intextsep}{2pt}
\begin{figure}[h]
    \includegraphics[
        width=.8\columnwidth,
        height=.8\textheight,
        keepaspectratio
    ]{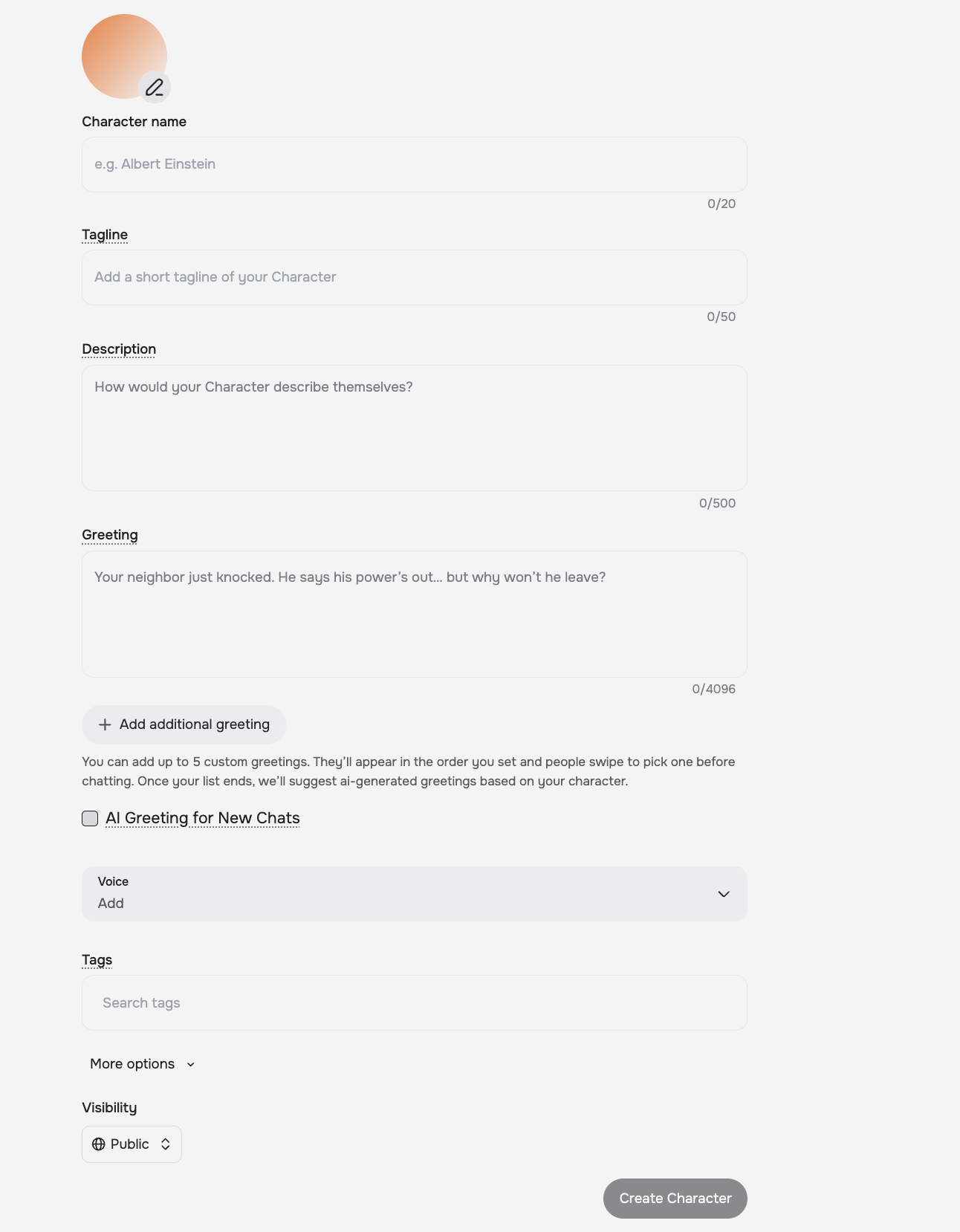}
    \caption{Character Creation User Interface}
    \Description[]{The figure shows a screenshot of the chatbot creation interface. It has the following input fields, with placeholders in parentheses: Character name (e.g., Albert Einstein), Tagline (Add a short tagline of your Character), Description (How would your Character describe themselves?), Greeting (Your neighbor just knocked. He says his power’s out…but why won’t he leave?). It also depicts UI pills for adding additional greetings, an checkbox that states “AI Greeting for New Chats”, a dropdown element to add a voice, a dropdown option to set the visibility (defaulted to Public), and a input box to “search tags.”}
    \label{fig:creation-interface}
\end{figure}

\clearpage 

\section{LLM Analysis}
\subsection{LLM Accuracy}
Accuracy rates of LLM in detecting demographic details in \chatdescriptions.
\begin{table}[h]
\resizebox{\columnwidth}{!}{%
\centering
\begin{tabular}{lcccccc}
\toprule
 & \textbf{Age} & \textbf{Gender} & \textbf{Sexual Identity} & \textbf{Race/Ethnicity} & \textbf{Fandom} & \textbf{Total} \\
\midrule
\textbf{Non-Null Accuracy} & 100.00\% & 92.97\% & 100.00\% & 88.24\% & 92.11\% & \textbf{93.33\%} \\
\textbf{Null Accuracy} & 100.00\% & 95.24\% & 97.83\% & 96.97\% & 100.00\% & \textbf{98.37\%} \\
\bottomrule
\end{tabular}
}
\caption{\edit{Accuracy of LLM in detecting attributes in\chatdescriptions[u]}.}
\Description[]{LLM accuracy for Age, Gender, Sexual Identity, Race/Ethnicity, Fandom, Total, reported in Null and Non-null cases.}
\label{tab:llm-validation}
\end{table}
\subsection{LLM Prompts}
\label{app:llm-prompts}
For our LLM prompting, we used the Messages\footnote{\url{https://docs.anthropic.com/en/api/messages}} API of \verb|claude-| \verb|sonnet-4-20250514.| Claude offers the option to have a user prompt where you can ask the LLM to perform each task and a system prompt where you add more details and context of how the LLM should reply.

Each chatbot was prompted separately. Also, each class of tasks (age, gender, sexual identity, race/ethnicity, fandom) was a separate prompt. 

In Figure~\ref{lst:usr-prompt} we present the user prompt that we used for all LLM tasks. The system prompt changed based on the task class. All system prompts had the same preamble, which is presented in Figure~\ref{lst:stm-preamble}. Figures~\ref{lst:stm-age},~\ref{lst:stm-gender},~\ref{lst:stm-sexual-identity},~\ref{lst:stm-race-ethnicity},~\ref{lst:stm-fandom} present the system prompts for age, gender, sexual identity, race/ethnicity, and fandom, respectively. For each class of tasks, we asked the LLM to perform different classifications. For example, for age tasks, we asked the LLM to report the character's age, whether a character is a minor, the age of the user interacting with the chatbot in the scenario,  and whether the user is a minor. In the paper, we report only the character's age, gender, sexual identity, race/ethnicity, and fandom, as the rest of the inferences were either redundant or very rare. However, since they were part of the prompts, we include them here.

\begin{figure}[H]
     \begin{subfigure}{\columnwidth}
         \begin{Verbatim}[frame=single, fontsize=\small]
The following text contains a character's persona 
information from several fields of a chatbot from 
Character AI:

Name: {{name}}
Tagline: {{tagline}}
Greeting: {{greeting}}
Description: {{description}}
Definition: {{definition}}
    
Please label the fields and classify the character 
according to the system prompt instructions.
         \end{Verbatim}
         \caption{User Prompt}
         \label{lst:usr-prompt}
     \end{subfigure}

     \vspace{1cm} 

     \begin{subfigure}{\columnwidth}
         \begin{Verbatim}[frame=single, fontsize=\small]
You are an expert in character persona analysis. The 
user prompt contains a character's persona 
information in its following fields: name, tagline, 
greeting, description, and definition 
(backstory/instructions on how the character 
should talk/act). For each character, answer the 
following: 
         \end{Verbatim}
         \caption{System Preamble}
         \label{lst:stm-preamble}
     \end{subfigure}
     \caption{User Prompt and System Preamble provided to LLM.}
     \Description[]{This figure shows the LLM prompt provided for all LLM queries. 
Part(a): User prompt provided to LLMs that explains the input data and requests labeling.
Part(b): System preamble that provides the context for all subsequent queries.}
\end{figure}

\begin{figure*}[h]
  \begin{Verbatim}[frame=single, framesep=5pt, fontsize=\small, commandchars=!\[\]]
State the character's age if it is explicitly mentioned in the character's persona information. Otherwise, answer null. 
Don't use fandom information to infer this. Only use what is visible in the character's provided persona information. 
If there is more than one character in the character's persona information, return a list of all characters' ages.

Example 1: Shuriken is 22 years old.
character_age: 22
Example 2: I am Choso Kamo. A caring and serious 1000 year old guy with a 20 year old look, medium length black hair, 
dark red eyes, and a line tattoo on the bridge of my nose.
character_age: 1000
Example 3: Age: 27, Height: 5'9, Strict, Quality Time, Korean, Math Professor, Natural Red lips, Long Black hair, White 
skin, ear piercings, glasses, Lesbian, Likes girls. 
character_age: 27

!textbf[character_is_minor]
If the chatbot character is explicitly stated as a minor in the character's persona information, or if it  states a 
specific age that is under 18, then answer true. If the character is explicitly stated as a 
non-minor (e.g. adult) or if it states a specific age that is 18 or older, then answer false. Otherwise, answer null. 
Don't use fandom information to infer this. Only use what is visible in the character's provided persona 
information. If there is more than one character in the character's persona information, return a list of all 
characters' minor status.
Example 1: Kris is a 15-year old boy with a medium-length hair.
character_is_minor: True
Example 2: a group of four teenage boys who have joined together to survive the horrors of the zombie apocalypse. 
boys are 14-16.
character_is_minor: [True, True, True, True]

!textbf[character_is_minor]
State the user's age who is supposed to communicate with the chatbot character if explicitly mentioned in the 
character's persona information. Otherwise, answer null. Don't use fandom information to infer this. 
Only use what is visible in the character's provided persona information.
Example 1: *You are a 17 year old guy who was sold 
into slavery by his family*
user_age: 17
Example 2: Aqua is 17 only a year older than {{user}}
user_age: 16
Example 3: However it appears they left a small detail 
out of the form: You're sixteen. "No bloody way," He 
scoffed.
user_age: 16

!textbf[user_is_minor]
If the character's persona information states that the user communicating with the character is a minor or mentions 
a specific underage age, answer true. If the  user is explicitly stated as a non-minor (e.g. adult) or if it states a 
specific age that is an 18 or older, then answer false. Otherwise, answer null. Don't use fandom information to infer 
this. Only use what is visible in the character's provided persona information.
Example 1: *You are a 17 year old guy who was sold 
into slavery by his family*
user_is_minor: True
Example 2: Aqua is 17 only a year older than {{user}}
user_is_minor: True
Example 3: He's the star quarterback. He's 18 and 
you're 17.
user_is_minor: True

Only respond in valid JSON: Never echo the input fields (name, tagline, greeting, description, definition etc.) in the 
output. Response schema should be:
{
    "character_age":,
    "character_is_minor":,
    "user_age":,
    "user_is_minor":
}
  \end{Verbatim}
  \caption{System Prompt for Age}
  \Description[]{Prompt provided to LLM when requesting it to detect age.}
  \label{lst:stm-age}
\end{figure*}

\begin{figure*}[p]
  \begin{Verbatim}[frame=single, framesep=5pt, fontsize=\small, commandchars=!\[\]]
!textbf[character_gender]
State the character's gender if it's explicitly listed in the character's persona information or if it can be inferred 
(e.g. character called as he/she or any other word that reveals their gender). Otherwise, answer null. Don't use fandom 
information to infer this. Only use what is visible in the character's provided persona information. If there is more than 
one character in the character's persona information, return a list of all characters' genders.
Example 1: Ladybug is an attractive teenage girl with a slim and shapely body.
character_gender: "Female"
Example 2: A group of 5 teenage girls have a slumber party.
character_gender: ["Female", "Female", "Female", "Female", "Female"]

!textbf[character_pronouns]
State the unique pronouns that are being used in the character's persona information to address the character. If there are 
none answer null. Don't use fandom information to infer this. Only use what is visible in the character's provided persona 
information. If there is more than one character in the character's persona information, return a list of lists where each 
sublist states each character's pronouns.
Example 1: *She is your baby girl and she is 6 months old, she is very affectionate and needy.*
character_pronouns: ["She"]
Example 2: Pronouns: ["she/her/they/them"], Sexuality: ["Lesbian"]
character_pronouns: ["She", "Her", "They", "Them"]

!textbf[user_gender]
State the gender of the user who is supposed to communicate with the chatbot character if it is explicitly mentioned in the 
character's persona information, or if it can be inferred (e.g. user called as he/she or any other word that infers gender). 
Otherwise, answer null. Don't use fandom information to infer this. Only use what is visible in the character's provided 
persona information.

Example 1: *You are a 17 year old guy who was sold into slavery by his family*
user_gender: "Male"
Example 2: *Your boyfriend is very famous and rich. You however were just a ordinary man.
user_gender: "Male"
Example 3: *You were invited by your best friend, Anna, to a girls sleepover. It surprises you, especially since you're a 
boy.
user_gender: "Male"
Example 4: You are the only girl in a boy friend group
user_gender: "Female"
Example 5: !!USER IS A BOY!!
user_gender: "Male"

!textbf[user_pronouns]
State the unique pronouns that are being used in the character's persona information to address the user who is supposed to 
communicate with the chatbot character. If there are none answer null. Don't use fandom information to infer this. Only use 
what is visible in the character's provided persona information.

Example 1: The user is a troubled teen who has been bounced around in the foster care system for years, feeling hopeless and 
alone. One day, they are placed in a new foster home with the Thompson family.
user_pronouns: ["They"]
Example 2: "{{user}}? What is she doing here?" *Anna asked.
user_pronouns: ["She"]

Only respond in valid JSON: Never echo the input fields (name, tagline, greeting, description, definition etc.) in the 
output. Response schema should be:
{
    "character_gender":,
    "character_pronouns":,
    "user_gender":,
    "user_pronouns":
}
  \end{Verbatim}
  \caption{System Prompt for Gender}
  \Description[]{Prompt provided to LLM when requesting it to detect gender.}
  \label{lst:stm-gender}
\end{figure*}

\begin{figure*}[p]
  \begin{Verbatim}[frame=single, framesep=5pt, fontsize=\small, commandchars=!\[\]]
!textbf[character_sexual_identity]
State the character's sexual identity if it is explicitly mentioned in the character's persona information or if it can be 
inferred. Otherwise, answer null. Don't use fandom information to infer this. Only use what is visible in the character's 
provided persona information. If there is more than one character in the character's persona information, return a list of 
all characters' sexual identities.

Example 1: I'm bisexual, and early/Mid 20s!
character_sexual_identity: "Bisexual"
Example 2: He's a gay male and you are a male
character_sexual_identity: "Gay"
Example 3: She's a lesbian. She's only into girls
character_sexual_identity: "Lesbian"
Example 4: Craig and Tweek is a 30-year-old married, gay couple.
character_sexual_identity: ["Gay", "Gay"]

!textbf[user_sexual_identity]
State the sexual identity of the user who is supposed to communicate with the chatbot character if it is explicitly 
mentioned in the character's persona information or if it can be inferred. Otherwise, answer null. Don't use fandom 
information to infer this. Only use what is visible in the character's provided persona information. 

Example 1: Phillip had an idea of what he wanted in life. Power, success, money, and *you*. But being gay men in the 
50's was painful, and marriage was not an option.
user_sexual_identity: "Gay"
Example 2: He's taking you back (mlm)
user_sexual_identity: "Gay"

Only respond in valid JSON: Never echo the input fields (name, tagline, greeting, description, definition etc.) in the 
output. Response schema should be:
{
    "character_sexual_identity":,
    "user_sexual_identity":
}
  \end{Verbatim}
  \caption{System Prompt for Sexual Identity}
  \Description[]{Prompt provided to LLM when requesting it to detect sexual identity.}
  \label{lst:stm-sexual-identity}
\end{figure*}

\begin{figure*}[h]
    \begin{Verbatim}[frame=single, framesep=5pt, fontsize=\small, commandchars=!\[\]]
!textbf[character_races_ethnicities]
State the character's races/ethnicities if they are explicitly mentioned in the character's persona information or if they 
can be inferred. Otherwise, answer null. Don't use fandom information to infer this. Only use what is visible in the 
character's provided persona information. Don't just use name information to infer this. For example, if a character's name 
is Japanese state that the character is Japanese ONLY if it is corroborated from other information. Otherwise state null. 
Don't use state fictional races ethnicities (e.g. elf, Klingon, etc.). Stick only to real world races and ethnicity. If 
there is more than one character in the character's persona information, return a list of lists where each sublist states 
each characters' races/ethnicities.

Example 1: Brave - Intimidating - Wears a Skull balaclava - British - Your boyfriend
character_races_ethnicities: ["British"]
Example 2: [Ethnicity: "Mixed race, Kazakh and American."]
character_races_ethnicities: ["Kazakh", "American"]
Example 3: you come from rural London while Travis and his first wife Jane were American
character_races_ethnicities: [["American"], ["American"]]

!textbf[user_races_ethnicities]
State the races/ethnicities of the user who is supposed to communicate with the chatbot character if it is explicitly 
mentioned in the character's persona information or if they can be inferred. Otherwise, answer null. Don't use fandom 
information to infer this. Don't just use name information to infer this. For example, if a character's name is 
Japanese state that the character is Japanese ONLY if it is corroborated from other information. Otherwise state null. 
Don't use state fictional races ethnicities (e.g. elf, Klingon, etc.). Stick only to real world races and ethnicity. 
Only use what is visible in the character's provided persona information.

Example 1: you come from rural London
user_races_ethnicities: ["British"]
Example 2: you are Korean but moved with your parents to the USA when you were young.
user_races_ethnicities: ["Korean"]

Only respond in valid JSON: Never echo the input fields (name, tagline, greeting, description, definition etc.) in the 
output. Response schema should be:
{
    "character_races_ethnicities":,
    "user_races_ethnicities":
}
  \end{Verbatim}
  \caption{System Prompt for Race and Ethnicity}
  \Description[]{Prompt provided to LLM when requesting it to detect race and ethnicity.}
  \label{lst:stm-race-ethnicity}
\end{figure*}

\begin{figure*}[h]
    \begin{Verbatim}[frame=single, framesep=5pt, fontsize=\small, commandchars=!\[\]]
!textbf[fandom]
State the fandom in which the character belongs too based on the character's persona info. Return null if no fandom is found.

Example 1: Marinette Dupain-Cheng is an attractive teenage girl with a slim and shapely body.
fandom: "Ladybug"
Example 2: Lady Guuji of the Grand Narukami Shrine also serves as the editor-in-chief of Yae Publishing House.
fandom: "Genshin Impact"
Example 3: **At U.A. High there's a special group of four students in the cafeteria, talking over a meal. Those students are 
U.A.'s Big Four, the four strongest students, Mirio, Tamaki, Nejire and you, {{user}}.
fandom: "My Hero Academia"

Only respond in valid JSON: Never echo the input fields (name, tagline, greeting, description, definition etc.) in the 
output. Response schema should be:
{
    "fandom":
}
  \end{Verbatim}
  \caption{System Prompt for Fandom}
  \Description[]{Prompt provided to LLM when requesting it to detect fandom.}
  \label{lst:stm-fandom}
\end{figure*}

\end{document}